\documentclass{emulateapj}

\slugcomment{Draft}

\shorttitle{IZI: Inferring Metallicities (Z) and Ionization Parameters
  (q)}
\shortauthors{Blanc et al.}

\begin{document}

\title{IZI: Inferring the Gas Phase Metallicity (Z) and Ionization
  Parameter (\lowercase{q})\\
 of Ionized Nebulae using Bayesian Statistics}

\author{Guillermo A. Blanc\altaffilmark{1}, Lisa
  Kewley\altaffilmark{2,3}, 
  Fr\'ed\'eric P.A. Vogt\altaffilmark{2}, Michael A. Dopita \altaffilmark{2,3,4}}

\altaffiltext{1}{Observatories of the Carnegie Institution for Science, Pasadena, CA, USA}
\altaffiltext{2}{Research School of Astronomy and Astrophysics, Australian National University, Cotter Rd., Weston ACT 2611, Australia }
\altaffiltext{3}{Institute for Astronomy, University of Hawaii, 2680 Woodlawn Drive, Honolulu, HI 96822, USA}
\altaffiltext{4}{Astronomy Department, King Abdulaziz University, P.O. Box 80203, Jeddah, Saudi Arabia}

\begin{abstract}

We present a new method for inferring the metallicity ($Z$) and
ionization parameter ($q$) of HII regions and star-forming galaxies using
strong nebular emission lines (SEL). We use Bayesian
inference to derive the joint and marginalized posterior probability
density functions for $Z$ and $q$ given a set of observed line fluxes and an input
photo-ionization model. Our approach allows the use
of arbitrary sets of SELs and the inclusion of flux upper limits. The method provides
a self-consistent way of determining the physical conditions of
ionized nebulae that is not tied to the arbitrary choice of a
particular SEL diagnostic and uses all the
available information. Unlike theoretically calibrated
SEL diagnostics the method is flexible and not tied to a particular
photo-ionization model. We describe our algorithm, validate it
against other methods, and present a tool that implements it
called IZI. Using a sample of nearby extra-galactic HII regions we assess the performance of 
commonly used SEL abundance diagnostics. 
We also use a sample of 22 local HII regions having both direct and recombination line (RL) oxygen
abundance measurements in the literature to study discrepancies
in the abundance scale between different methods.
We find that oxygen abundances derived through Bayesian
inference using currently available photo-ionization models in the literature
can be in good ($\sim$30\%) agreement with RL abundances, although some models
perform significantly better than others. We also confirm that abundances
measured using the direct method are typically $\sim0.2$ dex lower than
both RL and photo-ionization model based abundances.

\end{abstract}


\section{Introduction}

The ability to measure chemical abundances in the inter-stellar medium
(ISM) of galaxies is of fundamental
importance to astrophysics. We rely on these
measurements to constrain models of stellar evolution, galaxy
evolution, and the cosmological evolution of the universe as a whole.
The abundance of different elements in the ISM is modulated
by the processes driving stellar evolution (nucleosynthesis,
convection, winds, stellar rotation, mass ejection, SN explosions, etc.), as well as by the
way in which star formation proceeds both spatially and temporally in
the ISM of galaxies. Factors like the accretion and recycling of gas
from the inter-galactic (IGM) and circum-galactic (CGM) media, the
star formation efficiency (SFE) and star formation history (SFH), the initial
mass function (IMF), and the impact of feedback driven gaseous
outflows, all shape
the overall chemical structure of galaxies. The universe
therefore has a dynamical chemical history in which the abundance of heavy elements increases
as a function of cosmic time in a way that traces the efficiency with
which gas is turned into stars within galaxies and the way in which
these systems interact with the inter-galactic medium (IGM) by
accreting, enriching, recycling, and ejecting gas \citep[e.g.][]{larson74, tremonti04,
  dalcanton04, dalcanton07, brooks07, koppen07, finlator08,
  mannucci10, lara-lopez10, lara-lopez13, lilly13, zahid14}.

This article focuses on the measurement of chemical abundances in
nebulae ionized by recent star formation (i.e. HII regions). Ionized
gas has been widely used in the literature to trace the chemical composition of the
ISM. This is thanks to the fact that transitions from several common
elements are observable in emission at UV, optical and IR wavelengths and that the low
densities (10-10$^3$ cm$^{-3}$) typical of HII regions translate into
an optically thin medium for most of these transitions and a low level
of collisional de-excitation. These two conditions greatly simplify
the modeling of the emission spectra and allow a relatively
straightforward measurement of chemical abundances. 
There are three main methods used to measure
the metallicity or metal abundance\footnote{While the ideas
  exposed in this paper are in principle applicable to the chemical
  abundances of different elements, in this work we focus on the
  oxygen abundance (12+log(O/H)) which we assume traces the
  metallicity ($Z$). Section 2 provides a discussion on the
  relative abundance patterns used in this work.} in ionized gas which
we describe here briefly. For a review see \cite{stasinska04}.

\paragraph{The Direct Method:} The emissivity of collisionally excited
lines (CEL) depends strongly on the electron temperature ($T_e$). In this
method direct measurements of $T_e$ and the electron density ($n_e$) are used to
calculate the emissivity of CELs of
particular ions (e.g. [OII]$\lambda\lambda$3726,3729 or
[OIII]$\lambda\lambda$4959,5007 in the case of singly or doubly
ionized oxygen). Ionic abundances can then be
inferred from comparing the intensity of these lines to that of hydrogen
recombination lines (typically Balmer lines in the optical).  Assuming an
ionization correction to account for the abundance of unobserved
ions allows one to estimate the total elemental abundance. 
Auroral to nebular CEL temperature sensitive ratios are used to measure $T_e$
(e.g. [OIII]$\lambda$4363/[OIII]$\lambda$5007) while $n_e$ is
typically computed from the relative component intensity of density sensitive
doublets like [OII]$\lambda\lambda$3726,3729 and
[SII]$\lambda\lambda$6717,6731. The two main limitations of this method are first that
temperature sensitive auroral lines are too faint to be observed
directly in very distant and very high metallicity (i.e. low $T_e$) sources
($\sim10^{1-2}$ times fainter than H$\beta$), and second that the
method is sensitive to temperature fluctuations in nebulae which can
translate in underestimates of the abundance if a correction is not
applied \citep[e.g.][]{aller54, peimbert67,
peimbert69}. Furthermore, the existence of $T_e$ gradients in HII
regions implies that the direct method saturates for abundances higher
than solar \citep{stasinska02}.

\paragraph{The Recombination Line (RL) Method:} Unlike CELs, the emissivity
of RLs of ions show a very mild dependance on
$T_e$ and $n_e$ and their brightness relative to hydrogen
recombination lines scale practically directly with the ionic
abundance. An ionization correction for unobserved ions is still
necessary to obtain total elemental abundances using this
method. The RL method does not suffer from the biases
associated with temperature fluctuations that affect the direct method
and is though to be more robust \citep[although it might suffer from
its own set of systematic uncertainties][]{stasinska04}. On the other
hand, this same direct scaling with abundance translates into
extremely faint lines for elements heavier than He ($\sim10^{-4}$
fainter than H$\beta$ for O and C). Therefore RL abundance measurements require
extremely high $S/N$, high resolution spectra and have only been
made for He, C, and O in a few dozen bright HII regions in the Milky Way (MW) and the
Local Group \citep[e.g.][]{peimbert03, 
  peimbert05, tsamis03, esteban04, esteban09, garcia-rojas07,
  lopez-sanchez07, bresolin09}.

\paragraph{The Strong Emission Line (SEL) Method:} In order to be able
to measure abundances in faint, distant, and high metallicity nebulae
several authors have calibrated line ratios combining bright CELs and
Balmer lines (together called SELs) as abundance diagnostics. While
the brightness of SELs of different elements is strongly affected by
other parameters beyond the metal abundance, these calibrations take
advantage of the fact that in real HII regions correlations exist
between some of these parameters (e.g. $T_e$, the N/O abundance ratio,
the ionization parameter) and the metallicity. By exploiting these
``secondary'' correlations useful abundance diagnostic ratios can be
constructed out of SELs. 

Two different approaches have been taken in the literature to
calibrate SEL abundance diagnostics. The first approach is to
calibrate SEL ratios against direct method abundances using 
local samples of HII regions. These are typically referred to as
``empirical calibrations''. The second approach consists of
calibrating SEL ratios as a function of abundance using theoretical
photo-ionization models of HII regions. These are typically referred to
as ``theoretical calibrations'' .

Examples of empirical calibrations include those of the $R23=([OII]\lambda\lambda
3726,3729+[OIII]\lambda\lambda4959,5007)/H\beta$ ratio
\citep{pagel79, pilyugin05}, the $O3N2=[OIII]\lambda5007/[NII]\lambda6583$ ratio
\citep{alloin79, pettini04}, the $N2=[NII]\lambda6583/H\alpha$ ratio
\citep{denicolo02, pettini04}, the $S23=([SII]\lambda\lambda
6717,6731+[SIII]\lambda\lambda9069,9532)/H\beta$ ratio
\citep{vilchez96, diaz00}, and more sophisticated calibrations using
combinations of multiple line ratios like those presented in \cite{pilyugin11}
and \cite{pilyugin12}.

Theoretical calibrations use predictions
from full radiative transfer and excitation/ionization
calculations for idealized nebulae to calibrate the abundance
dependance of different diagnostics. Early works attempting this
include those by \cite{shields78, pagel79, dufour80,
  edmunds84, mccall85} and \cite{dopita86}. More recent theoretical
calibrations for individual diagnostics are given by \cite{mcgaugh91, kewley02} and
\cite{kobulnicky04} for the $R23$ ratio, and \cite{kewley02} for the
$O3N2$, $N2$, $S23$, and
$N2O2=[NII]\lambda\lambda6548,6583/[OII]\lambda\lambda3726,3729$
ratios. Even more recently \cite{dopita13} has calibrated several diagnostics
based on pairs of abundance and ionization sensitive ratios that can
be used to simultaneously constrain the metallicity and ionization
parameter of ionized nebulae \citep[see
also][]{kobulnicky04}.\\

The reliability of SEL methods to derive chemical abundances is
challenged by the alarming systematic differences seen between
different calibrations. These systematic biases include offsets of
0.2-0.6 dex in the absolute abundance scale derived using different
calibrations, non linearities (i.e. curvature) in the correlations between
results from different diagnostics, and dispersions of up to 0.3 dex
in abundance around these correlations. These issues have
been pointed out by several authors \citep{peimbert07, bresolin09,
  lopez-sanchez10, moustakas10}, and recently \cite{kewley08} and
\cite{lopez-sanchez12} have studied these biases in detail using a large
sample of galaxies from the Sloan Digital Sky Survey
\citep[SDSS,][]{york00} and theoretical HII region photoionization models
respectively. In these works a dominant trend arises in which
empirically calibrated SEL diagnostics yield
abundances that are systematically lower than those derived from
theoretically calibrated diagnostics.

Possible reasons behind these discrepancies have been thoroughly discussed in
the literature and \cite{lopez-sanchez12} provides an excellent
account. A problem with some SEL diagnostics is that they ignore the
dependance of the emission line ratios on the
ionization state of the gas. This is done by simply marginalizing over
the ionization parameter $q$ when calibrating a diagnostic using
photoionization models, inducing a large scatter and non-trivial
non-linearities in the method, or in the case of methods calibrated
against samples of HII regions with direct $T_e$ abundances by not
considering the dispersion and range in ionization parameter of the
calibration sample. A good example of the latter case are the $N2$ and
$O3N2$ calibrations of \cite{pettini04} which are widely used for high
redshift galaxies where the density, temperature, and ionization
conditions are most likely different from those in the HII
regions used to calibrate the method \citep[e.g.][]{brinchmann08,
  kewley13a, kewley13b, steidel14}.

Some authors have proposed SEL diagnostics that constraint the
ionization state of the gas at the same time of the chemical
abundance. These methods typically use recursive techniques or other
line ratios to compute either the ionization parameter $q$ \citep{mcgaugh91, kewley02,
  kobulnicky04, dopita13} or the empirically derived excitation parameter $P$
\citep{pilyugin05, pilyugin11}. Even when the ionization and
excitation state of the gas is considered, large systematic differences
are still seen between theoretically and empirically calibrated
methods. 

Another possible source of discrepancy is the potential
underestimation of the chemical abundance by the direct
method in the presence of density and temperature fluctuations and
temperature gradients in
ionized nebulae \citep{peimbert67, stasinska78, stasinska02}. HII
regions do not have homogeneous
density distributions, on the contrary, pockets, filaments, and shells
showing enhanced electron temperatures are ubiquitously observed in
these objects \citep[][and references within]{stasinska04}. In regions where $T_e$ is enhanced the
emissivity of the temperature sensitive auroral lines is
disproportionally enhanced over that of the stronger nebular lines. This
translates into an overestimation of the average $T_e$ and an
underestimation of the abundance when integrating the
line ratios over the whole nebula. While photoionization models assume
a uniform density distribution, they do include gradients in the
temperature distribution that somewhat alleviate this problem. A
related bias in the measurement of $T_e$ can also arise in the
presence of non-Maxwellian electron energy distributions
\citep[e.g. the $\kappa$ distribution introduced by][]{nicholls12,
  nicholls13}. 

Photo-ionization models on the other hand suffer form a series of
systematic uncertainties that are not well understood. For starters
they typically use libraries of synthetic stellar population models
to generate the input ionizing spectra for stellar populations of
particular metallicities, ages, and SFHs (see \S2). These models are
plagued by large uncertainties associated with the effects of
stellar rotation, stellar winds, and binarity, especially at the high
mass end of the IMF which is the most relevant for the ionization of
HII regions \citep[e.g.][]{meynet07, eldridge09, levesque10,
  steidel14}. Furthermore, photo-ionization models like the ones used
in this work typically assume a single-valued function describing the
nitrogen-to-oxygen abundance (N/O) as a function of
$12+\log{O/H}$. This is a reasonable assumption attempting to capture
the primary+secondary nature of nitrogen production but the observed relation
shows a large scatter and there is still disagreements regarding its
actual shape. Furthermore, the relation might change systematically
depending on the star formation, accretion, and gas ejection history
of galaxies. These assumptions regarding the N/O ratio can have
significant effects on diagnostics involving nitrogen lines
\citep{perez-montero09, perez-montero14, steidel14, belfiore14}.
Finally, assumptions regarding the ionization bounded nature of the
nebulae and their geometry can also influence the predicted line fluxes.

Given all these systematic uncertainties associated with current SEL
diagnostics, a more robust method that does not suffer from these biases
is pressingly needed. In this work we propose a technique to estimate the chemical abundance
and ionization parameter of HII regions which circumvents some, but
not all, of the problems discussed above by removing the need of ``calibrating'' a
particular SEL diagnostic, while instead doing a direct statistical
comparison between a theoretical photoionization model and all the
emission line data available to the observer. Our method uses the
formalism of Bayesian inference to derive the joint and marginalized
probability density distributions (PDF) of the parameters $Z$ and $q$
given the observed line fluxes and errors (the ``data''), a grid of
emission line ratios predicted by a theoretical photo-ionization model
(the ``model''), and any ``prior'' information on $Z$ and $q$. We also
provide and describe a publicly available user friendly {\tt IDL} implementation of our
method called {\tt IZI}\footnote{http://users.obs.carnegiescience.edu/gblancm/izi}
(Inferring metallicities (Z) and Ionization parameters). 

A similar approach to the one presented here was adopted by
\cite{brinchmann04} and \cite{tremonti04} to measure the physical
properties of star forming galaxies in the SDSS. More recently
\cite{perez-montero14} also presented a method based on $\chi^2$
minimization that uses information from multiple emission lines to
constraint the metallicity and ionization parameter. Also
worth mentioning is the pioneering work of \cite{garnett87} on a
sample of HII regions in M 81. This work was among the first to
explore the simultaneous use of multiple emission line ratios from
different elements to constraint the properties of HII regions by
comparing to photo-ionization model grids.

After presenting our new method (\S2) and the way in which it is implemented
in {\tt IZI} (\S3), we apply it to a sample of 186 extra-galactic HII
regions from \cite{vanzee98b} (\S4). We use this sample of HII regions
to validate our method against the results from the interpolation
method based on pairs of emission line diagnostics recently presented
in \cite{dopita13}. 

The Bayesian formalism provides us with a tool to evaluate how much
information regarding the metallicity and the ionization parameter is
carried by different subsets of the data (i.e. by different emission
lines). This provides an opportunity to evaluate the intrinsic
performance of different diagnostics in terms of
constraining these two parameters. By intrinsic performance we mean
the base-line amount of information carried by a particular line
ratio, independent of any biases in its calibration. We evaluate the
performance of a series of commonly used SEL diagnostics in the
literature and present these results in \S5. 

We also apply {\tt IZI} to a sample of local HII regions having both
direct method and RL abundance measurements in order to explore the
abundance scale discrepancies discussed above. Our method allows us to
perform this comparison using several input photo-ionization models
available in the literature. The RL abundances provide an important
reference point from which to obtain insight regarding the discrepancy
between direct method and empirical SEL abundances and those derived
from photo-ionization models. We present the results of this comparison
in \S6. Finally we summarize our work and provide our conclusions in \S7.

\section{The Method: Bayesian Inference of $Z$ and $\lowercase{q}$}

In this section we describe the statistical method used to derive the parameters
$Z$ and $q$ from a set of observed SEL fluxes. In the context of
Bayesian statistics this is a ``parameter estimation'' problem in
which we want to estimate the joint and marginalized PDF for the
parameters $\vec{\theta}=\{{\rm log}(Z),{\rm log}(q)\}$ given the data $D=\{\vec{f},
\vec{e}\}$ (where $\vec{f}=\{f_1\dots
f_n\}$ are the observed line fluxes and $\vec{e}=\{e_1\dots
e_n\}$ their associated measurement errors), the model $M$ given by
a photoionization model grid of line fluxes $\vec{f'}(\vec{\theta})=\{f'_0(\vec{\theta})\dots
f'_n(\vec{\theta})\}$, and any prior information $I$ regarding the parameter
values. We follow the notation of \cite{gregory05}. 

From Bayes theorem it can be shown that the joint posterior
PDF of $\vec{\theta}$ is given by 

\begin{equation}
p(\vec{\theta}\;|D,M,I)=\frac{p(\vec{\theta}\;|M,I)\;p(D\;|M,\vec{\theta},I)}{p(D\;|M,I)}
\end{equation}

For simplicity, following the notation of \cite{gregory05} we
omit $I$ in the following equations, although it is understood that all
conditional probabilities are calculated given the prior information on the
parameter values. We can then rewrite Equation 1 as

\begin{equation}
p(\vec{\theta}\;|D,M)=\frac{p(\vec{\theta}\;|M)\;p(D\;|M,\vec{\theta})}{p(D\;|M)}
\end{equation}
 
The first term in the numerator of the right hand side reflects the
prior probability on the parameters $\vec{\theta}$ given the model $M$
and all available prior information on the parameters. The second term
in the numerator of the right hand side of Equation 2
is the ``likelihood'' of the data given the model evaluated for
parameters $\vec{\theta}$. Assuming
Gaussian errors in the flux this term is given by

\begin{equation}
p(D\;|M,\vec{\theta})=\frac{1}{2\pi}\prod_{1}^n
\frac{\exp(-\frac{(f_i-f'_i(\vec{\theta}))^2}{2(e_i^2+\epsilon^2
    f'_i(\vec{\theta})^2)})}{\sqrt{e_i^2+\epsilon^2 f'_i(\vec{\theta})^2}}
\end{equation}

\noindent
where $n$ is the number of observed emission lines and $\epsilon$ is a
term that takes into account the systematic
fractional error in the flux predicted by the photo-ionization model. As a
zeroth order approximation we assume that this systematic error is
Gaussian and is the same for all modeled emission lines. For the results
presented in this paper we adopt a fractional error
of 0.1 dex \citep{kewley08, dopita13}. In principle $\epsilon$
could be turned into an extra ``nuisance'' parameter in the
calculation. In sake of computing speed and simplicity we do not adopt this approach,
although we expect to experiment with this in the future.
The denominator in the right hand side of Equation 2 is the ``global
likelihood'' of the data for model $M$. This is equivalent to a
normalization factor and can be calculated by
demanding the integral of $p(\vec{\theta}\;|D,M)$ to be
unity. 

Once the joint posterior PDF for $\vec{\theta}$ is known, we can
compute the marginalized posterior PDF for $\log (Z)$ and $\log (q)$ by
integrating the joint posterior PDF over the other parameter respectively, so 

\begin{equation}
p(\;\log (Z)\;|D,M)=\int_{q_{min}}^{q_{max}}
p(\vec{\theta}\;|D,M)\; d\log (q)
\end{equation}

\begin{equation}
p(\;\log (q)\;|D,M)=\int_{Z_{min}}^{Z_{max}}
p(\vec{\theta}\;|D,M)\; d\log (Z)
\end{equation}

The joint and marginalized posterior PDFs can be used to derive
``best-fit'' values and confidence limits for $Z$ and $q$ by either
searching for the maximum in the distributions (equivalent to a
maximum likelihood approach) or by computing
the moments of the distributions. In some cases the shape of the PDF
will be such that a best-fit value will not be a well defined
quantity. This can happen when the data
only provide upper or lower limits on a parameter, in which case
the PDF is only bounded on one side, or in cases in which the PDF
shows a complex topology with multiple probability peaks reflecting
multiple solutions\footnote{A common example of this situation occurs
  when using the lines that compose the $R23$ diagnostic, which is known
to be double valued in metallicity.}. In the next section we describe how the
method outlined above is implemented in {\tt IZI} and how we deal with
these sort of complications. 

This Bayesian approach has
six major advantages over classical SEL methods to
derive chemical abundances and ionization parameters in ionized
nebulae:

\begin{enumerate}

\item This method removes the arbitrary choice of a particular SEL diagnostic
    and it therefore does not suffer from many of the systematic
    uncertainties associated with the calibration of these diagnostics.

\item The method can make use of all the information available to the
    observer, as an arbitrary set of emission lines can be used as
    long as they have been modeled as part of the photo-ionization
    grid.

\item The method allows for the inclusion of upper limits in emission line
  fluxes as the Bayesian approach allows us to include censored data
  in the likelihood calculation in a straightforward manner. This is
  particularly useful when studying the properties of high redshift
  galaxies where upper limits in the line fluxes are common.

\item {\tt IZI} is implemented in a flexible fashion that allows the
    use of user provided photo-ionization model grids. Therefore,
    unlike SEL theoretical calibrations in the literature, the method
    is not ``married'' to a particular set of models. This implies
    that {\tt IZI} can be run using different input photo-ionization
    models in order to test how the results change as a function of
    the underlying assumptions in the models.

\item In certain cases the PDFs of $Z$ and $q$ can have complicated
    topologies, showing multiple local maxima and asymmetric
    probability tails that reflect non Gaussian uncertainties in the
    derived values. Since the method does a full calculation of the
    joint and marginalized PDFs it naturally yields information about
    the asymmetry in the error bars of the derived parameters, the
    existence of multiple degenerate solutions, and on whether a
    parameter is being constrained or only a limit can be set on it.

\item {\tt IZI} is fast, easy to use and to incorporate as a
    subroutine into larger pieces of code, and is publicly available
    for use by the astronomical community.

\end{enumerate}

In the following sections we describe how this method is implemented
and applied to different samples of HII regions.

\section{IZI: A User Friendly IDL Implementation}

We have implemented the method described in \S2 as an {\tt IDL} function
called {\tt IZI}. Here we provide a detailed description of how this function
computes the chemical abundance and ionization parameter from a set of
input emission line fluxes and a photo-ionization model grid.

\subsection{Input SEL Fluxes}

Input fluxes and errors are assumed to be corrected for dust
extinction. All fluxes are normalized to the flux of the H$\beta$ line
or, if the flux of this line is not provided, to the flux of the brightest line
for which a measurement is available. Upper limits on the fluxes of
particular lines can be provided and the algorithm deals with them by
integrating the argument of the product in Equation 3 below these limits. Besides the measured line fluxes and
errors the user must also provide an array identifying the name of each line. These
IDs are also stored in the input photo-ionization
model grids and are used to properly match the data and the
models. {\tt IZI} allows the user to define sets of
lines for which the flux must be co-added before fitting to allow for
low spectral resolution measurements that do not separate the flux of
neighboring transitions.

\subsection{Input Model Grids} 

Our method is flexible in the sense that it can use an arbitrary
model grid as long as the observed emission lines have
been modeled. This is an advantage of {\tt IZI} as, unlike theoretically
calibrated SEL diagnostics, the method is not married to a particular
set of photo-ionization models. We stress that any arbitrary model can be
provided as input to {\tt IZI} as long as it is stored in the
appropriate format. In principle this is not restricted to
photo-ionization models but could also include empirically based grids
of emission line fluxes as a function of abundance and
ionization/excitation parameter, similar to the approach used in the
C-method of \cite{pilyugin12}. Although we have not experimented with
this possibility this would be an interesting application for {\tt
  IZI}. Here we briefly describe the photo-ionization models used
throughout this paper. We refer the reader to the original references
for a more detailed description.

\subsubsection{\cite{kewley01} Models}

\cite{kewley01} present a suite of photo-ionization models aimed at
reproducing the emission line properties of a sample of starburst galaxies in the
local universe. The authors use the MAPPINGS-III photo-ionization code \citep{sutherland93}
to compute the nebular emission spectra of idealized HII regions
ionized by synthetic stellar populations. \cite{kewley01} use both the
PEGASE v2.0 \citep{fioc97} and STARBURST 99 \citep{leitherer99}
stellar population synthesis codes to compute the input ionizing
spectra. Here we only adopt the STARBURST 99 models which use the
plane-parallel stellar atmosphere models of \cite{lejeune97} plus
the extended model atmospheres of \cite{schmutz92} for stars with
strong winds in combination with the Geneva stellar evolution tracks
\citep{schaller92}. The input spectra are computed
assuming a constant star formation history, a Salpeter IMF \citep{salpeter55} with
a lower mass cutoff of 0.1 $M_{\odot}$ and an upper mass cutoff of 120
$M_{\odot}$, and an age of 8 Myr which allows for the stellar birth
and death rate to balance for stellar masses $>20$ M$_{\odot}$ and for
Wolf-Rayet (W-R) star emission to develop and contribute to the
ionizing spectrum. 

The synthetic spectra ionize MAPPINGS-III isobaric
plane-parallel photo-ionization models with electron densities of $n_e=10$ cm$^{-3}$
and 350 cm$^{-3}$. In these models the ionization
parameter\footnote{Sometimes reported in the literature in its
  dimensionless form ${\cal U}=q/c$ with $c$ the speed of light.}
is defined at the inner boundary of the nebular models by the following
expresion:

\begin{equation}
q=\frac{Q_{H^0}}{n_H}
\end{equation} 

where $Q_{H^0}$ is the flux of ionizing photons above the Lyman limit
and $n_H$ is the particle density. 
For all elements except He and N the MAPPINGS-III calculations adopt
an undepleted solar abundance pattern and assume primary production. For
He a primordial nucleosynthesis component is added and the N abundance
follows a broken power-law as a function of metallicity in an attempt
to account for secondary N production above $0.23Z_{\odot}$. The
MAPPINGS-III code includes the effects of dust absorption, dust photoelectric
heating, and element depletion as described in \cite{dopita00}. 

\subsubsection{\cite{levesque10} and \cite{richardson13} Models}

\cite{levesque10} presents new MAPPINGS-III photo-ionization models
aimed at reproducing the nebular spectra of large samples of galaxies
in the local universe. This new version of the MAPPINGS-III models is
similar in nature to the \cite{kewley01} models described above but
include a series of upgrades with respect to the older
models. \cite{levesque10} adopt an updated version of the STARBURST 99 population
synthesis code \citep{vazquez05} that uses the Pauldrach/Hillier stellar atmosphere models
\citep{pauldrach01, hillier98}. These models, unlike the
Lejeune/Schmutz models used in \cite{kewley01}, include a detailed
non-LTE modeling of metal line blanketing which significantly affect
the shape of the ionizing spectrum making it harder in the 1-4 Ry
region. \cite{levesque10} used two updated versions of the Geneva
tracks with ``standard'' and ``high'' mass loss rates \citep{meynet94}. Here we
adopt the models calculated using the ``high'' mass loss tracks, a
constant star formation history, a Salpeter IMF with a 100 $M_{\odot}$
upper cutoff, and an age of 6 Myr. 

These synthetic spectra are used to ionize an updated version of the
MAPPINGS-III photo-ionization models that includes a more rigorous
treatment of dust that includes the effects of radiation pressure on
grains \citep{groves04}. \cite{levesque10} also adopt isobaric
plane-parallel models but with electron densities of $n_e=10$ cm$^{-3}$
and 100 cm$^{-3}$. The ionization parameter and elemental abundance
patterns are defined in the same way as in the \cite{kewley01}
models. \cite{richardson13} presents an extension of these models that
samples very high values for the ionization parameter. These models, which
are aimed at reproducing the emission line properties of high redshift
Ly$\alpha$ emitters, are computed following the same prescriptions as
in \cite{levesque10} and therefore can be merged with these grids in a
straightforward manner.

\subsubsection{\cite{dopita13} Models}

Recently \cite{dopita13} presented photo-ionization model grids
computed with a significantly upgraded version of MAPPINGS. The
MAPPINGS-IV code has been updated to include the
latest atomic data, an increased number of ionic species treated as
full non-LTE multi-level atoms, and the ability to use electron energy
distributions that differ from a simple Maxwell-Boltzmann (M-B)
distribution. \cite{dopita13} adopt a $\kappa$ electron energy
distribution that shows a power-law shaped high energy tail. This distribution is inspired by observations
of plasmas in the Solar System and its inclusion is found to solve
many discrepancies in the measurement of the electron temperatures in
planetary nebulae and HII regions \citep{nicholls12, nicholls13}. The grids used
here are calculated by adopting a $\kappa$ electron energy
distribution with values of  $\kappa=20$ and $\kappa=\infty$ (the
latter being equivalent to a M-B distribution). The new version of the
code also uses a revised solar abundance set from \cite{grevesse10}
and new smooth functions to parametrize the abundance of both N and C
as a function of O abundance to account for secondary production of
these two elements,

The input ionizing spectrum is computed using the STARBURST 99
population synthesis code adopting the Lejeune/Schmutz extended
stellar atmosphere models \citep[as in][]{kewley01}, a constant star
formation history, a Salpeter IMF with a lower mass cutoff of 0.1
$M_{\odot}$ and an upper mass cutoff of 120 $M_{\odot}$, and an age of
4 Myr. Unlike the \cite{kewley01} and \cite{levesque10} models,
\cite{dopita13} computes isobaric photo-ionization models with an
electron density $n_e\simeq10$ cm$^{-3}$ and a spherical geometry
for which the radial divergence of the radiation field must be taken
into account when computing the ionization parameter. For these
spherical models the ionization parameter is defined at the inner
boundary of the nebulae by:

\begin{equation}
q=\frac{Q_{H^0}}{4\pi R_s^2 n_H}\propto(Q_{H^0}\; n_H)^{1/3}
\end{equation}

where $R_s$ is the Str\"omgren radius of the spherical HII region
\citep{stromgren39}. \\

\subsection{Priors and Limits on $log(Z)$ and $log(q)$}

By default we assume a maximum ignorance situation and use a uniform
PDF to compute the first term in the numerator of the right hand side
of Equation 2. Notice that since we use the logarithm of $Z$ and $q$ as
the model parameters this is equivalent to using a Jeffreys prior on
both parameters, implying that we assume uniform probability per
decade in the parameters $Z$ and $q$
\citep{gregory05}. We do this instead of adopting a uniform prior in
linear space since the latter assigns an unreasonably high probability
to high abundance and high ionization parameter solutions. {\tt IZI}
also allows the user to provide an arbitrary prior PDF for each
parameter independently. For example one could introduce a prior
for $Z$ based on the stellar mass of a star-forming galaxy and the
observed scatter in the $M-Z$ relation. We assume that
the parameters live within a range ${Z_{min},Z_{max}}$ and ${q_{min},q_{max}}$
which by default are taken as the minimum and maximum values modeled in the
input photo-ionization grid, but that can also be set by the user. The
prior probability is set to zero outside this range. The implications
of this last assumption for the calculation of upper and lower limits
in $Z$ and $q$ are discussed in \S3.5. 

\subsection{Grid Interpolation and Calculation of the PDFs}

We conduct a full calculation of the joint posterior PDF over a finely spaced
grid in both $\log (Z)$ and $\log (q)$. Since the input
photo-ionization grids typically sample a coarse set of values for these
parameters, before computing the PDF we use either a bi-linear or a quintic surface
interpolation to resample the line flux values in the input model onto
a finely spaced grid. By default the grid has $2500=50\times50$ equally spaced
elements in the $\log (Z)$-$\log (q)$ plane, within the
${Z_{min},Z_{max}}$ and ${q_{min},q_{max}}$ ranges. {\tt IZI}  allows
for the number of elements to be set by the user but computing time is
strongly affected by this choice.

Once the input model has been interpolated we use Equations 2 and 3 to
calculate the joint posterior PDF at each element of the parameters
grid. The marginalized posterior PDF for each parameter is calculated
then using Equations 4 and 5.

\subsection{Calculations of Best-fit Values and Confidence Intervals}

Choosing how to define a ``best-fit'' value for a parameter is non
trivial when its PDF is non-Gaussian. Strong asymmetries and the
presence of multiple peaks (i.e. multiple likely solutions) in the PDF
can deviate the mean of the PDF from the most likely value (mode),
making the mean not representative of a likely solution for
the parameter in question. This is of course alleviated by computing
the full PDF which in principle removes the need to calculate a
best-fit value in the first place. In practice however, for many
applications it is not straightforward to propagate the full PDF of each
parameter through all subsequent calculations in the analysis, and a
having a best-fit value complemented by a confidence interval is
desirable. Here we describe how we calculate the best-fit values
for $Z$ and $q$, although we remind the reader that a proper
analysis should make use of the full PDFs \citep[e.g.][]{brinchmann04}.

We calculate three different best-fit values for $\log (Z)$ and
$\log (q)$ respectively. The first is the ``marginalized mean'',
given by

\begin{equation}
\langle log (Z) \rangle=\int_{Z_{min}}^{Z_{max}}
p(\;\log (Z)\;|M,\vec{\theta})\log (Z)\; d\log (Z)
\end{equation}

\begin{equation}
\langle log (q) \rangle=\int_{q_{min}}^{q_{max}}
p(\;\log (q)\;|M,\vec{\theta})\log (q)\; d\log (q)
\end{equation}

\noindent
The other two are the modes of the joint posterior PDF and of the
marginalized posterior PDFs. That is, the values of
$\log (Z)$ and $\log (q)$ that maximize either the joint PDF or the
marginalized PDF of each parameter independently. We refer to these
from hereon on as the ``joint mode'' and ``marginalized mode''.

We calculate a single confidence interval for each parameter by defining
the high probability region in the $\log (Z)$-$\log (q)$ plane that
encloses 68.3\% of the integrated joint posterior probability
density. The parameter values that bound
this region are taken as our 1$\sigma$ uncertainty on the best-fit
values. Using this definition we find that the confidence interval
always bounds the three ``best-fit'' values defined above.

\subsection{Identifying Problematic PDFs}

{\tt IZI} automatically analyzes the shape of the marginalized PDFs in
order to flag PDFs that are not ``well behaved'' in the sense that
the best-fit values discussed above might not be representative of the
actual parameter values. We use the first and second derivatives of
the marginalized PDFs to search for maxima, and we characterize the
number of peaks in the distributions, flagging objects that show
multiple possible solutions in $Z$ and $q$. In these cases
the marginalized mean is not ensured to fall close to a region of high
probability, and the joint and marginalized modes might not
sample the highest probability solution as a wider lower amplitude peak
in the PDF might be actually more likely than a narrower higher
amplitude peak (i.e. the mode) once the probability density
is integrated over them.

We also study the PDFs to check if the solutions are bounded or only
provide upper or lower limits or no limits at all on $Z$ and $q$. A
solution is considered bound if the probability density at ends of the
parameter range (${Z_{min},Z_{max}}$ and ${q_{min},q_{max}}$) is lower
than 50\% of the marginalized mode. Otherwise the object is flagged as
showing an upper limit, lower limit, or no limit at all on the
corresponding parameter.

\section{Application to Real Sources and Validation of the Method}

In this section we apply the method described in the last two sections
to a sample of 186 extragalactic HII regions observed and catalogued
by \cite{vanzee98b} (hereafter V98). We also compare the results of
{\tt IZI} against measurements of the metallicity and ionization
parameter for these same objects using the methods described in
\cite{dopita13}. The goal of this section is to test the performance of
{\tt IZI} on real data and validate our method. 

\subsection{Application to Observed Local HII Regions}

\begin{figure*}[t]
\begin{center}
\epsscale{1.0}
\plotone{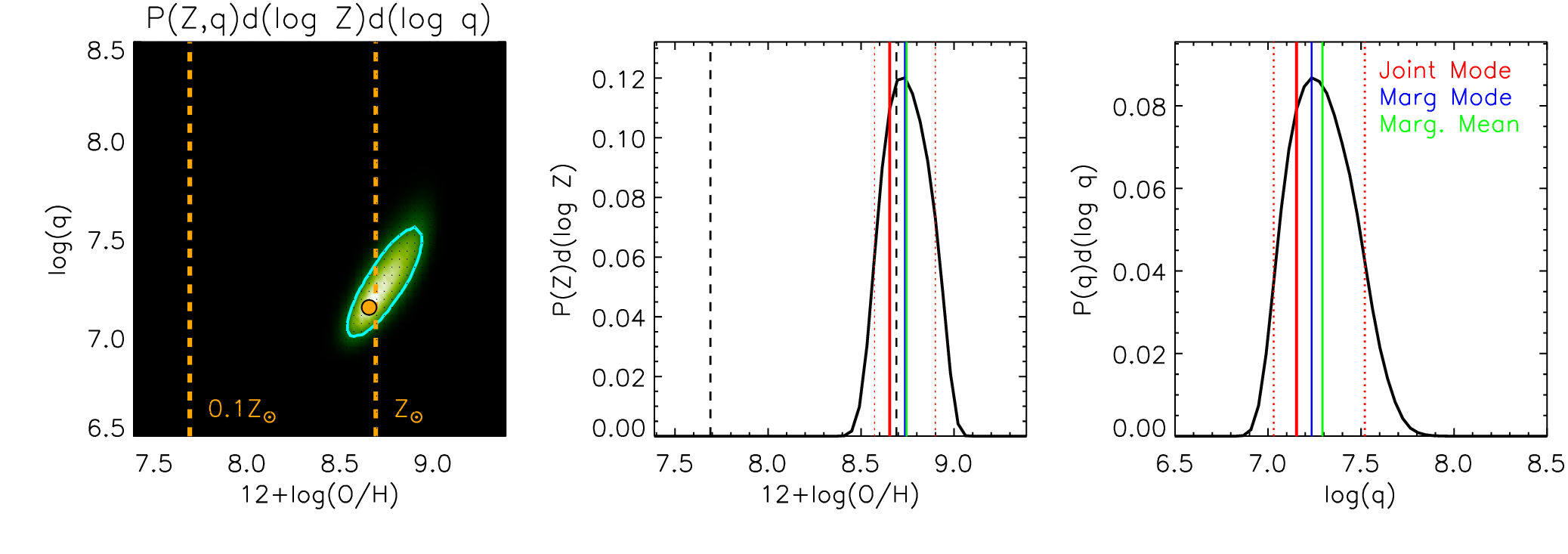}
\plotone{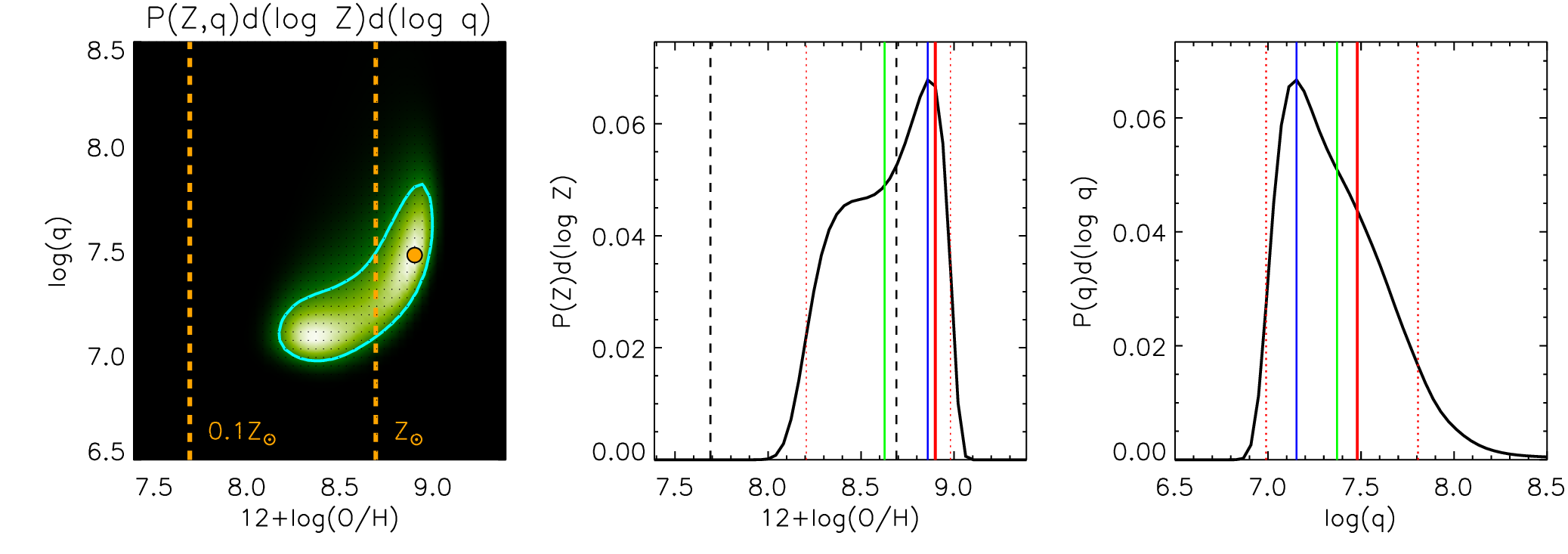}
\plotone{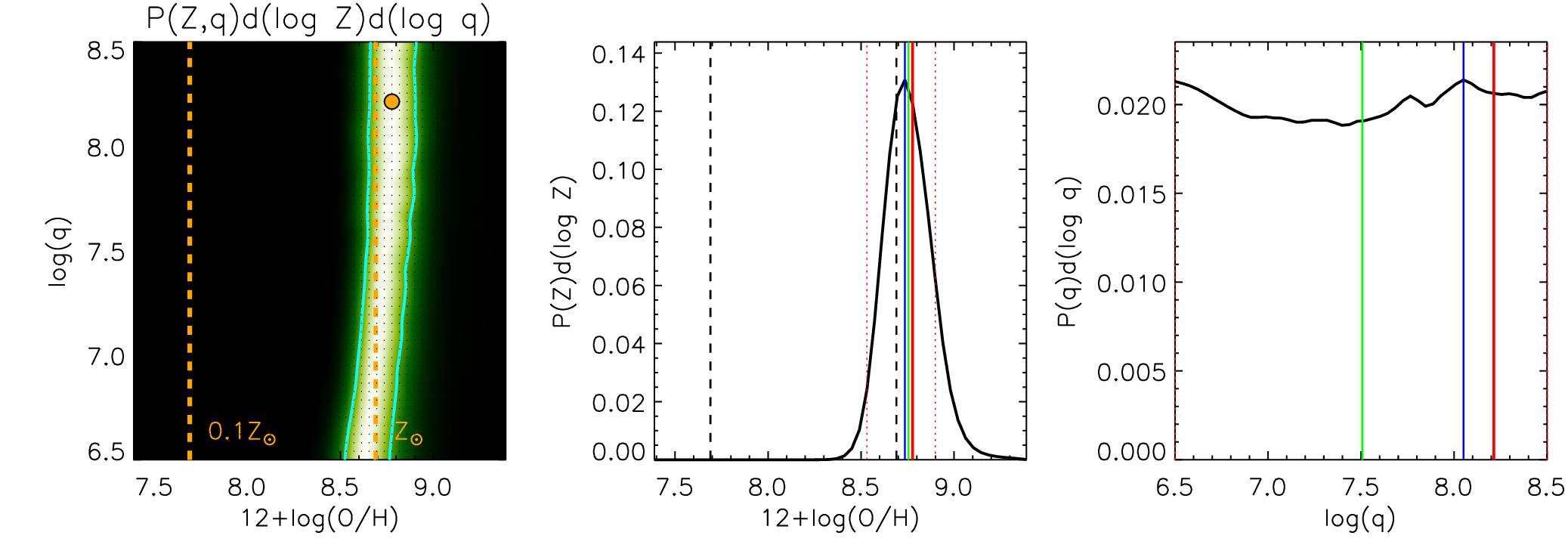}
\plotone{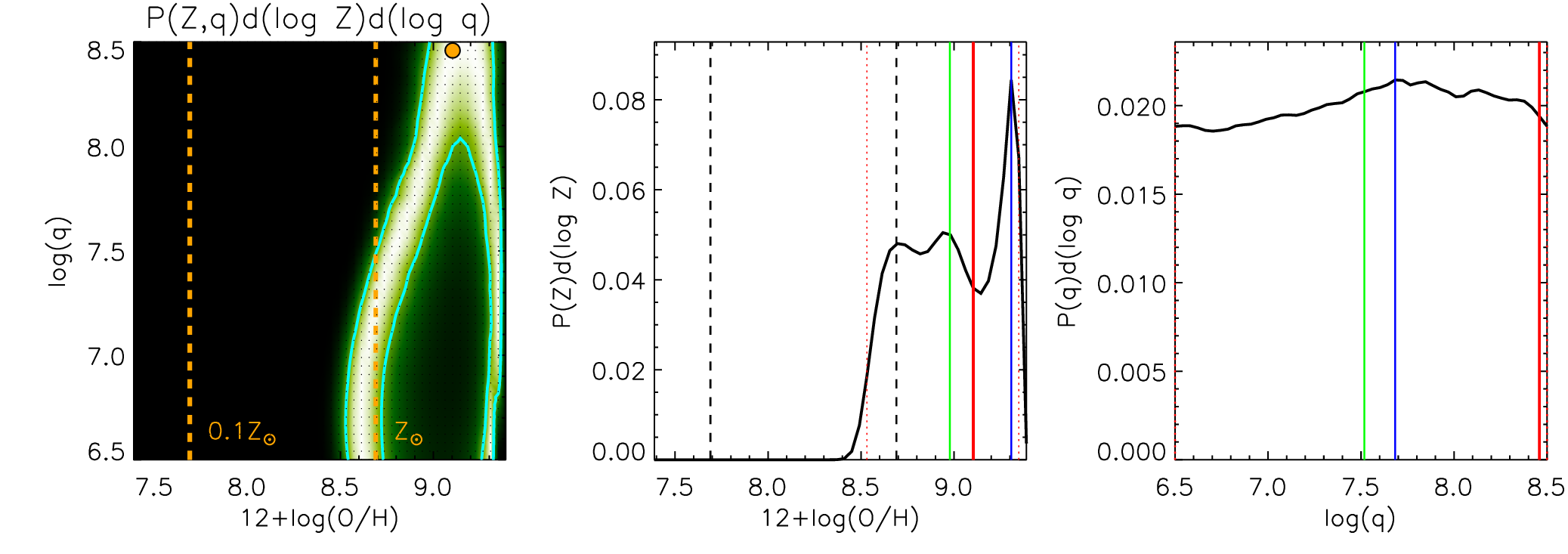}
\caption{{\it First row:} Joint (left) and marginalized (center,
  right) PDFs for $\log{Z}$ and $\log{q}$ calculated by {\tt IZI} for
  the HII region NGC 0925 +087-031 using all emission line measurements reported
  in V98 (i.e. [OII]$\lambda\lambda$3726,3729, H$\beta$,
  [OIII]$\lambda$5007, H$\alpha$, [NII]$\lambda$6583, and
  [SII]$\lambda\lambda$6717,6731). The orange circle and cyan line in
  the left panel show the mode of the joint PDF and the 1$\sigma$
  confidence level. The dashed orange (black) lines in the left
  (middle) panel mark the adopted solar
  abundance and a tenth of its value. On the middle and right panels
  the joint mode, marginalized mode, marginalized mean, and 1$\sigma$
  confidence interval are shown by solid red, blue, green, and dashed
  red vertical lines respectively. {\it Second row: } Same as
  above, but only using the [OII]$\lambda\lambda$3726,3729, H$\beta$,
  and [OIII]$\lambda$5007 lines (i.e. equivalent to $R23$).  {\it Third row:} Same as above
  but only using the [OII]$\lambda\lambda$3726,3729 and
  [NII]$\lambda$6583 lines (i.e. equivalent to $N2O2$).  {\it Bottom row:} Same as above
  but only using the H$\alpha$ and
  [NII]$\lambda$6583 lines (i.e. equivalent to $N2$).}
\label{fig-1}
\end{center}
\end{figure*}

We use the data of V98 who uniformly obtained low
resolution ($\Delta\lambda=7.8$\AA) spectra for 186
extragalactic HII regions in a sample of 13 nearby spiral galaxies
using the Double Spectrograph on the \mbox{5 m Palomar} telescope. Their data
covers the 3500-7600\AA\ range and therefore includes the
[OII]$\lambda\lambda$3726,3729, H$\beta$,
[OIII]$\lambda\lambda$5007, H$\alpha$,
[NII]$\lambda\lambda$6583, and [SII]$\lambda\lambda$6717,6731
lines that we use here\footnote{While V98 reports the summed
  intensity of the [OIII] and [NII] doublets, here we assume
  the theoretical line ratio to compute the flux of
  [OIII]$\lambda$5007 and [NII]$\lambda$6583. This is done for
  consistency with \cite{dopita13} in order to allow the comparison
  presented in \S4.2}. The publicly available catalog reports
reddening corrected emission line fluxes (normalized to H$\beta$) and errors for all the above
transitions. This sample was also used by \cite{dopita13} to evaluate
the performance of their newly proposed metallicity and excitation
diagnostics, so it provides a good reference for comparing the
results of {\tt IZI} to that work. In order to allow for this
comparison in this section we use the $\kappa=20$ photo-ionization models of
\cite{dopita13} as input to {\tt IZI}.

The top panels of Figure \ref{fig-1} present the joint and marginalized PDFs calculated by
{\tt IZI} using all the above SELs for a single HII region in the V98 sample (NGC
925 +087-031). For this particular object both $Z$ and $q$ are well
constrained and show positive covariance. The marginalized PDF for
$\log{(Z)}$ is roughly Gaussian while the $\log{(q)}$ PDF shows some
asymmetry towards large values. The three ``best-fit'' values for
the two parameters in the model are shown by the red (joint mode),
blue (marginalized mode), and green (marginalized mean)
solid lines in the two rightmost panels. The confidence intervals
described in \S3.5 are shown by the cyan curve in the joint PDF and
the dotted red lines in the marginalized PDFs. For the metallicity it
is seen that all three estimations agree well within the 1$\sigma$
confidence limits. The joint mode is $12+\log{O/H}=8.70^{+0.22}_{-0.15}$, close to
the solar value. For the ionization parameter the joint mode is
$\log{q}=7.19^{+0.37}_{-0.16}$, and the marginalized mean is offset to
a slightly higher value because of the asymmetry in the PDF, although
all three estimations fall well within the 1$\sigma$ confidence limits.

To exemplify the behavior of the PDF as a function of the available
input information, the second row of panels in Figure \ref{fig-1} shows the joint and
marginalized PDFs now calculated using only the
[OII]$\lambda\lambda$3726,3729, H$\beta$ and
[OIII]$\lambda\lambda$5007 transitions. This is similar to using the
$R23$ diagnostic. The double valued nature of
$R23$ as a metallicity diagnostic is caused by the strengthening of the
oxygen lines with respect to H$\beta$ as the abundance increases,
followed by a weakening of these lines towards higher abundances
($12+\log{O/H}\gtrsim 8.5$) as a higher metal cooling rate causes
the temperature, and therefore the rate of collisional excitations to
the upper levels of these oxygen transitions to decrease. This degeneracy has been
extensively documented in the literature \citep[e.g.][]{pagel79,
  kobulnicky04, kewley08} and arises naturally when using our method
as two high probability peaks seen in the joint PDF. 

\begin{figure*}[t]
\begin{center}
\epsscale{1.1}
\plotone{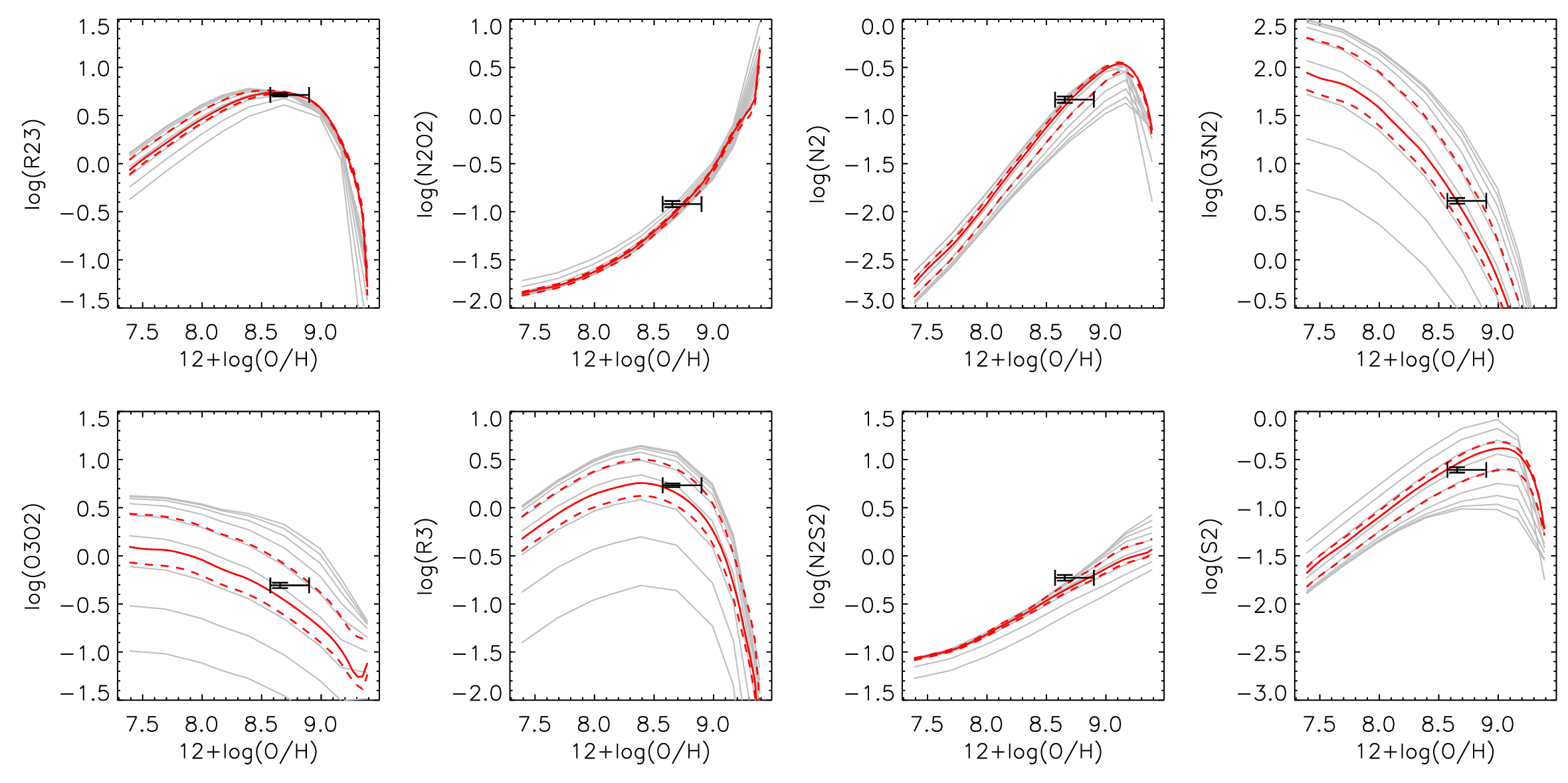}
\rule{0.8\textwidth}{.4pt}
\plotone{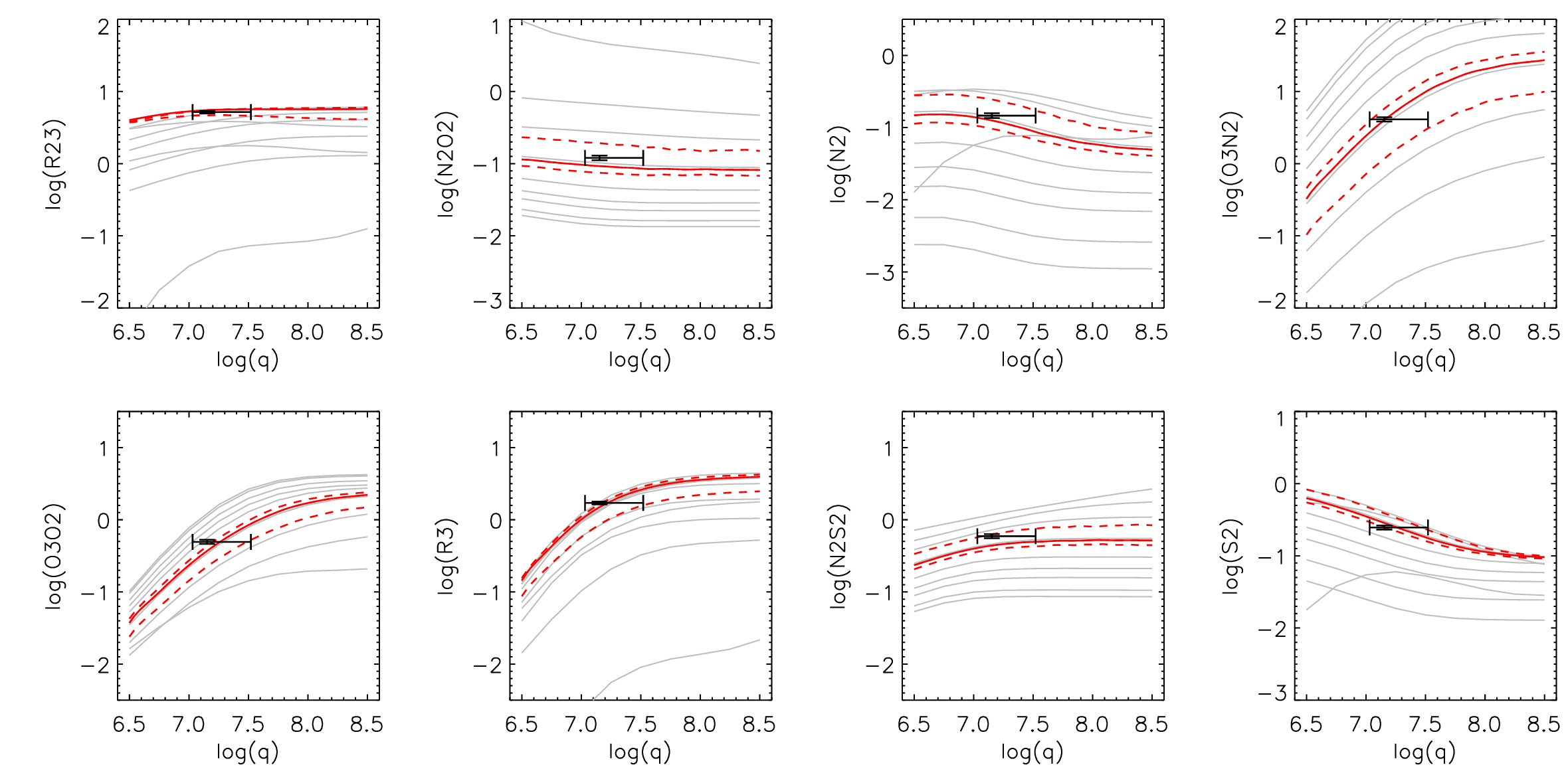}
\caption{Eight line ratios commonly used in the literature as metallicity and excitation
diagnostics ($R23$=([OII]$\lambda\lambda$3726,3729+[OIII]$\lambda$5007)/H$\beta$;
  $N2O2$=[NII]$\lambda$6583/[OII]$\lambda\lambda$3726,3729;
  $N2$=[NII]$\lambda$6583/H$\alpha$;
  $O3N2$=[OIII]$\lambda$5007/[NII]$\lambda$6583;
  $O3O2$=[OIII]$\lambda$5007/[OII]$\lambda\lambda$3726,3729;
  $R3$=[OIII]$\lambda$5007/H$\beta$;
  $N2S2$=[NII]$\lambda$6583/[SII]$\lambda\lambda$6717,6731; and $S2$=[SII]$\lambda\lambda$6717,6731/H$\alpha$), as a function of oxygen abundance
  (top eight panels) and ionization
parameter (bottom eight panels). The solid gray lines show the input
photo-ionization model grids \citep{dopita13} for different values of $q$ in the top eight
panels and different values of $Z$ in the bottom six panels. The
best-fit solution (joint mode) for \mbox{NGC 0925 +087-031} is shown by the
solid red line and the black error-bars, and the families of allowed
photo-ionization models within the 1$\sigma$ confidence limits are
bounded by the dashed red lines. }
\label{fig-2}
\end{center}
\end{figure*}

In this case, since {\tt IZI} uses the individual line fluxes for the
input transitions and not the sum of the singly and doubly ionized
oxygen lines (as in the $R23$ diagnostic), it also constrains the
ionization parameter although it shows a broader and more asymmetric PDF
than when all transitions are used. In this sense, using {\tt IZI}
with these transitions as input is somewhat similar to the
$R23$ method of \cite{kobulnicky04} in which the [OIII] to [OII] ratio
is used in combination with $R23$ to constrain the ionization parameter and
the metallicity simultaneously using a recursive method.

The third row of panels in Figure \ref{fig-1} shows the PDFs calculated by
{\tt IZI} when using only the [OII]$\lambda\lambda$3726,3729 and
[NII]$\lambda$6583 lines (i.e. equivalent to the $N2O2$
diagnostic). The similar ionization potentials of N$^+$ and O$^+$
translate into this line ratio being very insensitive to the
ionization parameter as can be seen in Figure \ref{fig-1} where $q$
is absolutely unconstrained. On the other hand, both the secondary nature
of N production and the fact that the energy of the upper level of the
[NII] doublet is lower than that of the upper level for [OII]
doublet (making its line flux less sensitive to a drop in $T_e$ towards
higher abundances) translate in this line ratio being very sensitive to
metallicity \citep{dopita00, kewley02}. This can be seen
in the PDFs shown in Figure \ref{fig-1} where the metallicity is
well constrained while the ionization parameter is completely
unconstrained. It is worth mentioning that $N2O2$ is very sensitive to
the N/O abundance ratio \citep{perez-montero09, perez-montero14}, and
that its good performance as a metallicity diagnostic relies heavily on
the assumption of a correlation between N/O and O/H in the
photo-ionization models being used here. We discuss this further in \S5.

Finally, the bottom row of panels in Figure \ref{fig-1} shows the
resultant PDFs if only the [NII]$\lambda$6583 and H$\alpha$ lines are
fed to {\tt IZI} as input (i.e. equivalent to the $N2$ diagnostic). In
this case the ionization parameter is completely unconstrained, and
the abundance show solutions for two branches (see Figure 2 and
discussion in \S5.2) [NII]$\lambda$6583. The topology of the joint PDF
translates into three peaks in the marginalized PDF for the
abundance. In each of these branches the
best-fit abundance is highly correlated with the ionization
parameter. It is interesting to note that if the ionization parameter
is constrained to the value derived when using all available lines the
correct value for the abundance is recovered in the low abundance
branch. This reflects the fact that the performance of $N2$ as an abundance
diagnostic relies heavily on the fact that the ionization parameter of
local HII regions is limited to a fairly narrow range and that the
large majority of them is in the low abundance branch of $N2$. We will discuss
this further in \S5.2.

In order to visualize where the constraining information on the
parameters $Z$ and $q$ is coming from, Figure \ref{fig-2} presents eight
line ratios commonly used in the literature as metallicity and excitation
diagnostics, as a function of oxygen abundance and ionization
parameter. The solid gray lines show the input
photo-ionization grids for different values of $q$ in the top eight
panels and different values of $Z$ in the bottom eight panels. The
best-fit solution (joint mode) for NGC 0925 +087-031 (using all
available lines) is shown by the
solid red line and the black error-bars, and the families of allowed
photo-ionization models within the 1$\sigma$ confidence limits are
bounded by the dashed red lines. All panels showing dependance
against $Z$ ($q$) sample a common range of 3 dex (4 dex) in the line ratios and
therefore provide a fair picture of how sensitive these ratios are to
these parameters.

Many insightful conclusions can be drawn from Figure 2. First, the
$R23$, $N2O2$, $N2$, and $N2S2$ diagnostics are better metallicity indicators than
the $O3N2$, $O3O2$, $R3$, and $S2$ ratios. The former show less
dependance on the ionization parameter, which translates into a small
scatter as a function of metallicity. The $R23$, $N2$, and $N2S2$
diagnostics show mild dependances with $q$ in different regions of the
abundances range which translates into larger scatter for these
diagnostics in these regions. The $N2$ diagnostic also shows a
flattening towards the high abundance end that is caused by the same
process that flattens and inverts the behavior of$ R23$ (i.e. a drop in
$T_e$ towards high abundances), so this diagnostic is also doubled
valued (see discussion in \S5.3.2). The $N2O2$ diagnostic on the other
hand is monotonic over the whole abundance range and shows the least
dependance on $q$ and therefore the smallest scatter as a function of
metallicity. These two factors make it a good abundance
diagnostic, modulo variations in the N/O ratio which can be significant.

The $O3N2$, $O3O2$, $R3$, and $S2$ line ratios show a strong dependance on the
ionization parameter. In fact, $O3O2$ has long been recognized
as a good excitation diagnostic \citep[e.g.][]{mcgaugh91, lilly03,
  nakajima13}, and out of all the line ratios shown in Figure \ref{fig-2} it shows
the smallest dependance on the metallicity. Nevertheless, these line
ratios are sometimes used in the literature as metallicity diagnostics,
empirically calibrated against local samples of HII regions
\citep[e.g.][]{nagao06, maiolino08, mannucci09, cullen13}. The reason why
these empirical calibrations appear to trace metallicity with a
reasonably moderate scatter is that the local samples against which
they are calibrated only span a restricted range in ionization
parameters. Therefore, extrapolating the use of these calibrations to
other objects that might have different excitation conditions
(e.g. high redshift galaxies) can introduce significant biases in
the derived abundances \citep{kewley13a, steidel14}. This is a possible reason behind the
discrepancies seen among different authors studying the behavior of
the mass-metallicity-SFR relation as a function of redshift
\citep[e.g.][]{mannucci10, troncoso13, cullen13}.

Figure 2 also exemplifies how {\tt IZI} allows the information from
all available emission lines (and upper limits if available) to be used
in order to constrain the values of $Z$ and $q$. This is a major
improvement over classical metallicity and excitation diagnostics
which are typically limited to either a single SEL ratio or a pair of
SEL ratios. 

\subsection{Comparison to pyqz}

In order to validate the output of {\tt IZI} we compare our
results for the full V98 sample against those produced by the
publicly available Python module {\tt pyqz}\footnote{{\tt pyqz} was
  developed at ANU by Fred\'eric Vogt and it is publicly available
  at: http://dx.doi.org/10.4225/13/516366F6F24ED} presented in
\cite{dopita13}. The {\tt pyqz} code uses pairs of abundance and
excitation sensitive line ratios to define a plane in which the
metallicity and ionization parameter can be determined by
interpolating the photo-ionization model grid to match the observed
line ratios. 

\begin{figure}[t]
\begin{center}
\epsscale{1.0}
\plotone{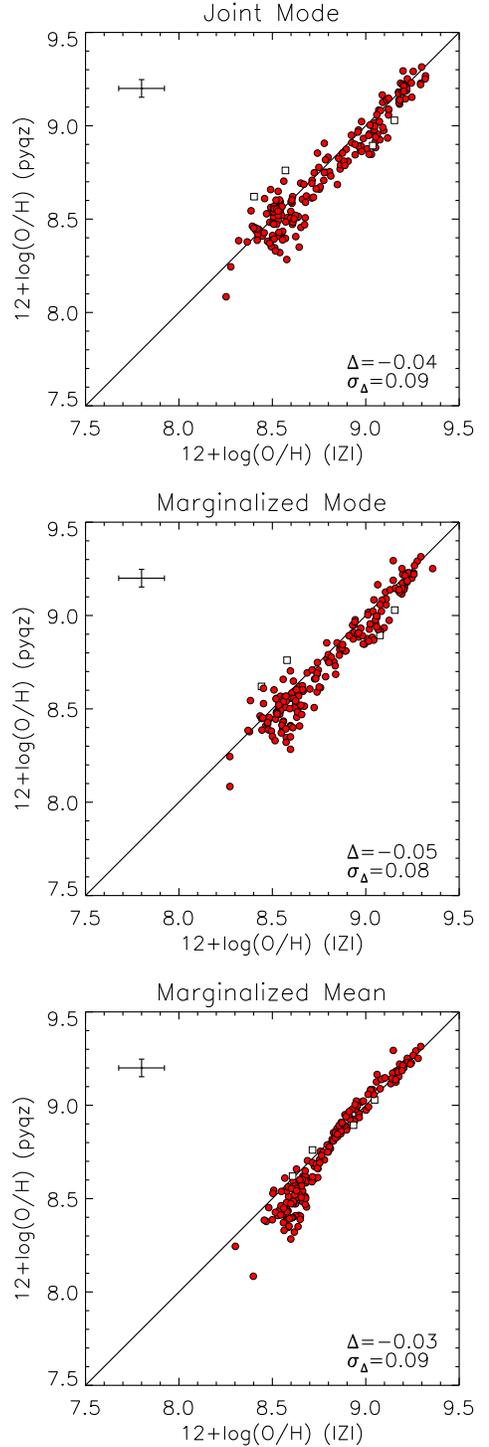}
\caption{Comparison between the joint mode (top), marginalized mode
  (middle) and marginalized mean (bottom) best-fit values for the
  metallicity estimated with {\tt IZI} for the V98 sample using all
  available emission line fluxes, and the average of eight {\tt pyqz}
  estimations using different diagnostic pairs of SELs (see
  text). Objects shown in red have ``well behaved'' posterior PDFs for the
  metallicity, meaning that they are not flagged by {\tt IZI} as
  having either upper or lower limits in $Z$, nor multiple peaks in the
  PDF. Squares show objects with multiple peaks in the marginalized
  metallicity PDF, while circles show single peak objects. Median
  error-bars for the full sample are shown.}
\label{fig-3}
\end{center}
\end{figure}

\begin{figure}[t]
\begin{center}
\epsscale{1.0}
\plotone{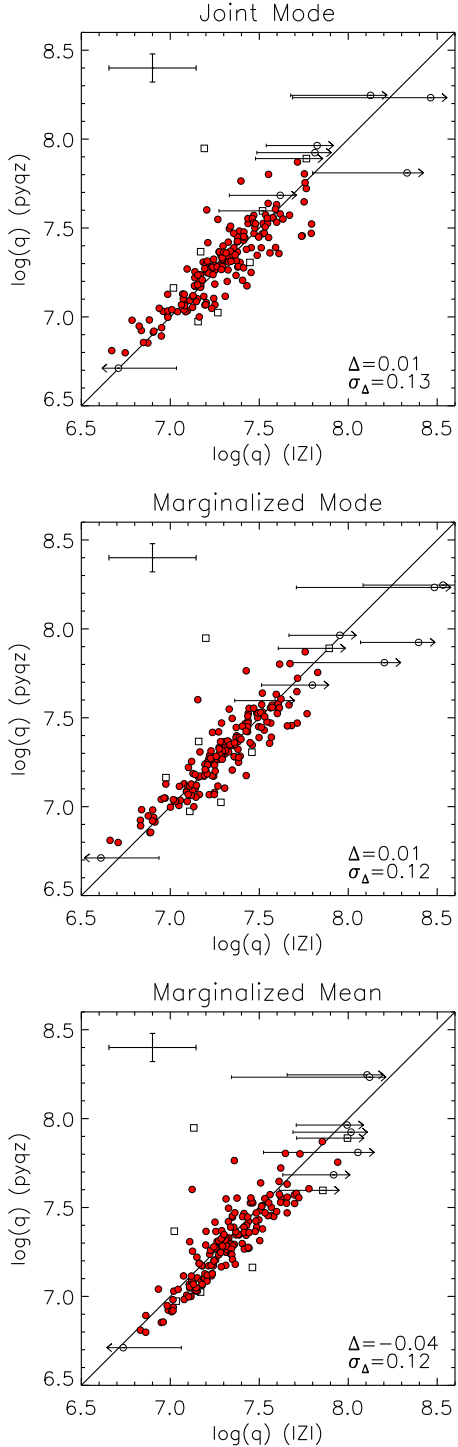}
\caption{Same as Figure \ref{fig-3} but for the ionization parameter $q$.}
\label{fig-4}
\end{center}
\end{figure}

\cite{dopita13} shows that parameter estimations using their
new updated photo-ionization model grids and {\tt
  pyqz} on four different pairs of line ratios
based on combinations of the [NII]/[OII] and [NII/SII] abundance
sensitive ratios and the [OIII]/[OII], [OIII]/[SII], and
[OIII]/H$\beta$ excitation sensitive ratios, yield self-consistent
results. Figures \ref{fig-3} and \ref{fig-4} present a comparison between the results of
{\tt IZI} for the full V98 HII region sample and the
average of eight {\tt pyqz} metallicity and ionization parameter estimations
based on the following pairs of line ratios:

\begin{itemize}
\item{} [NII]/[OII] vs [OIII]/[OII]
\item{} [NII]/[OII] vs [OIII]/[SII]
\item{} [NII]/[SII] vs [OIII]/H$\beta$
\item{} [NII]/[SII] vs [OIII]/[SII]
\item{} [NII]/[OII] vs [OIII]/H$\beta$
\item{} [NII]/[SII] vs [OIII]/[OII]
\item{} [NII]/H$\alpha$ vs [OIII]/H$\beta$
\item{} [NII]/H$\alpha$ vs [OIII]/[OII]
\end{itemize}

The {\tt IZI} values are calculated using all the emission line fluxes
reported in V98 and the \cite{dopita13} model grids for
$\kappa=20$. The exact same information is used in the {\tt
  pyqz} calculations so any differences between the two
methods arise from the different approach taken to estimate the $Z$
 and $q$ parameters. That is, the difference between the average of
 eight different interpolations using different pairs of abundance and
 excitation sensitive ratios, and a full posterior PDFs calculation
 followed by a best-fit parameter estimation.

Figure \ref{fig-3} shows good agreement between the ``joint mode''
best-fit abundance (top panel) as calculated by {\tt IZI} and the average of the
eight {\tt pyqz} estimations. The measurements agree well within the
{\tt IZI} 1$\sigma$ confidence limits with a median offset of -0.03
and a scatter of 0.08 dex. Note that the median error-bar
reported in Figures \ref{fig-3} and \ref{fig-4} for the {\tt pyqz}
measurements correspond to the standard deviation between the eight
measurements and do not include neither measurement errors nor
systematic uncertainties, so they are surely underestimated.

The marginalized mode (second panel of Figure \ref{fig-3}) shows
similar agreement with the average {\tt pyqz} values, but significant systematic deviations at
the 0.2 dex level can be seen when using the marginalized mean (bottom
panel), particularly at the low abundance end
($12+\log{O/H}<8.7$). This is caused by asymmetries in the
marginalized PDF which shift the mean away from the highest
probability value. 

The ionization parameter also shows good agreement between the two
methods within the 1$\sigma$ confidence limits as can be seen in
Figure \ref{fig-4}. In this case we do not see significant differences
between our three best-fit estimates of $q$. Overall, the consistency
with the \cite{dopita13} methods as implemented in {\tt pyqz} is
excellent. This is reassuring and implies that our Bayesian inference
method is able to recover the metallicity and ionization parameter of
local HII regions as well, if not better, than the state of the art
methods currently available in the literature, while also providing
all the advantages listed in \S1.  

Hereafter we adopt the joint mode as the best estimate of the $Z$ and
$q$ parameters. This is because it incorporates information regarding
the covariance between the two parameters (unlike the marginalized
mode) and because it is not dependent on how asymmetric the PDF is
(unlike the marginalized mean).

\section{Evaluating the Performance of Individual SEL Diagnostics}

In Figures \ref{fig-1} and \ref{fig-2} we exemplified how the resulting PDFs depend on the
information carried by different subsets of emission lines for a
single HII region in the V98 catalog. Here we extend this analysis
to the full V98 sample in order to statistically evaluate the
performance of different SEL diagnostics that have been proposed in
the literature and discuss the reasons behind the systematic
differences observed. Having access to the actual shape of the
metallicity and ionization parameter PDFs shows to be revealing in
terms of understanding the sources of systematic biases associated with
different diagnostics. 

By adopting the best-fit abundances derived using
the full set of emission lines available for the V98 sample as
fiducial ``true'' values, and comparing them to estimates of the
abundance done using subsets of these lines (i.e. emulating different
SEL diagnostics) we can study the origin of the
scatter associated with different methods, as well as the causes
behind any systematic deviations from the ``true'' values.

In this section we use {\tt IZI} to evaluate the performance of the
four diagnostics recently proposed by \cite{dopita13} as well as a set
of six classic diagnostics commonly used in the literature: $R23$,
$N2O2$, $N2$, $O3N2$, $O3O2$, and $R3$. These line ratios are
commonly used to estimate the
metallicity of HII regions and star forming galaxies in the local
universe as well as at high redshift. All these diagnostics have been
calibrated against local samples of HII regions, theoretical
photo-ionization models, or in some cases combinations of the two
\citep[e.g.][]{maiolino08}. Here we are not interested in studying
the differences between these calibrations, but rather we want to understand how sensitive different
diagnostics are to the ionized gas metallicity, and what are the
intrinsic biases and expected scatter associated with them. To this
effect we do not use the original calibrations proposed in the
literature and instead we run {\tt IZI} using only the subset of SELs
associated with each diagnostic and the \cite{dopita13} grids as the
input model. The comparison between these results and
the fiducial metallicities computed using all available SELs allows us
to evaluate the intrinsic performance of these diagnostics in terms of
how much information they carry regarding the metal abundance
independently of the way in which they are calibrated. In \S6 we
revisit the subject of abundance discrepancies caused by different
types of calibrations.

\subsection{\cite{dopita13} Diagnostics}

\cite{dopita13} propose a set of new diagnostics
consisting of four pairs of SEL ratios able to simultaneously constrain the abundance
and ionization parameter of nebulae. These correspond to the first four
methods listed in \S4.2. Figure \ref{fig-5} presents a comparison between
the fiducial metallicities derived with {\tt IZI} and the results of using
only the lines associated with these four diagnostics. Below we
comment on the performance of each:

\begin{figure*}[t]
\begin{center}
\epsscale{1.1}
\plotone{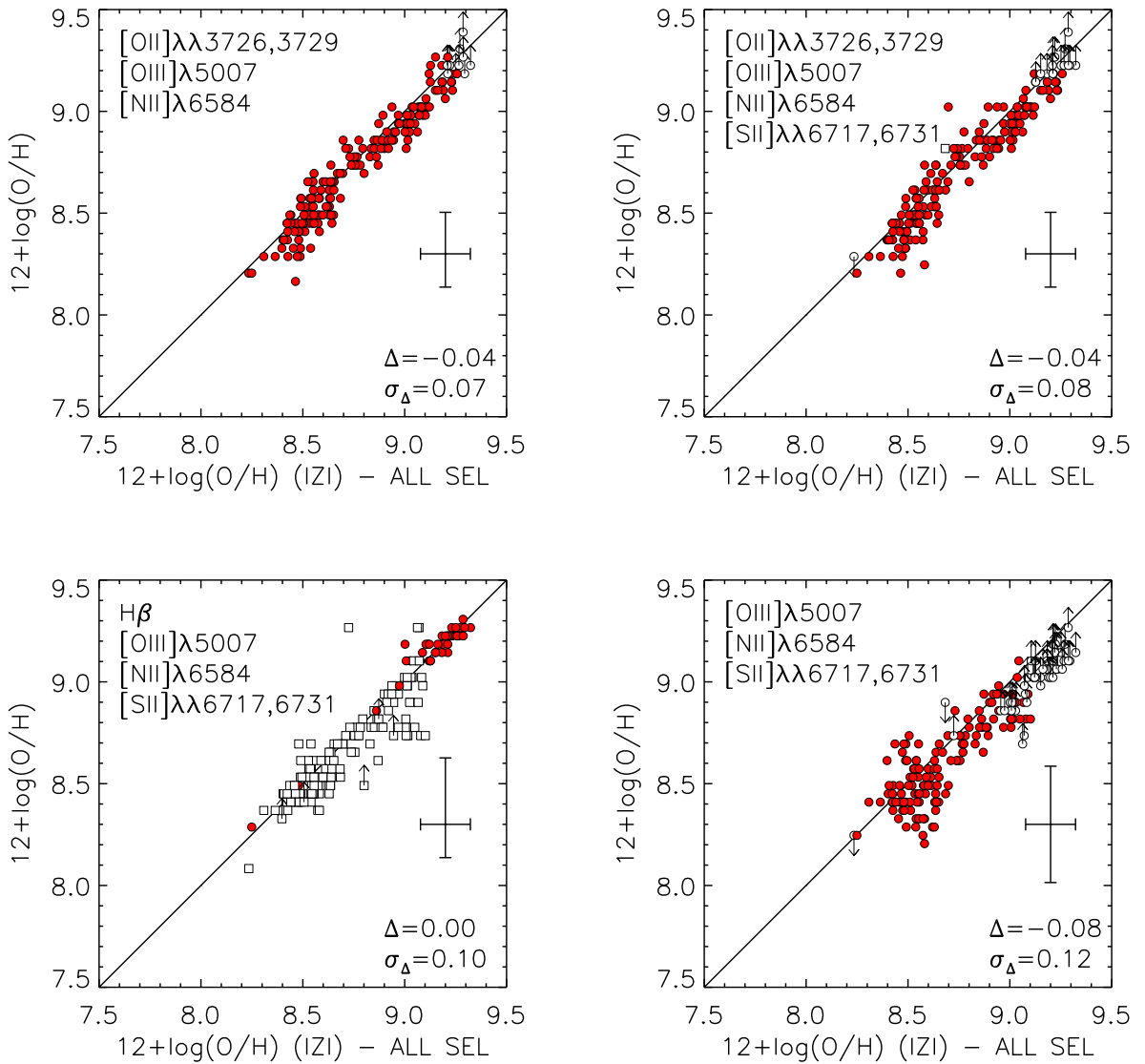}
\caption{Comparison of {\tt IZI} derived abundances using all lines
  available in the V98 catalog (x-axis) versus abundances calculated
  using subsets of these lines (y-axis). In each plot reports the
  subset of lines being used to compute the abundance in the y-axis,
  the median offset and the standard deviation of the datapoints, and
  the median error-bars for all points. Circles show objects with
  single peaked marginalized abundance PDFs while squares show objects
 with multiple peaks in the PDF. Upper and lower limits are marked by
 arrows and indicate objects in which the PDF is not bounded on one
 side. Objects having satisfying both criteria (i.e. single-peaked
 and bounded PDF) are shown in red.}
\label{fig-5}
\end{center}
\end{figure*}

\paragraph{[NII]/[OII] vs [OIII]/[OII]} This diagnostic (top-left
panel of Figure \ref{fig-5}) performs well over the whole range of
sampled metallicities with a median offset of -0.04 dex and a scatter of
0.07 dex. At the high abundance end ($12+\log{O/H}>9.2$) {\tt IZI} only provides lower limits on the
metallicity when using this restricted set of lines. This is because
the posterior PDF becomes wider when only using a subset of the lines
and its high abundance end starts to overlap with the $Z_{max}$
value (i.e. the highest abundance modeled in the input
photo-ionization model grids). The limits are therefore
associated with the limits of the input model grids and do not reflect
any physical process taking place at these metallicities.

\paragraph{[NII]/[OII] vs [OIII]/[SII]} This diagnostic (top-right
panel of Figure \ref{fig-5}) also performs well with a median offset of -0.04 dex and a
scatter of 0.08 dex. It shows the same issue regarding the lower
limits in the metallicity at the high abundance end as the previous
diagnostic. Besides objects having lower limits in $Z$, one HII region
(NGC 4395 +099-029) at $12+\log{O/H}\simeq8.7$ is flagged as having
multiple peaks in the marginalized abundance PDF. Close inspection
shows that this object presents a particularly elevated [SII] flux,
which is in tension with the best-fit model preferred by the [OII],
[OIII], and [NII] lines. This gives rise to a secondary low abundance
(i.e. low [NII]/[SII] ratio) peak in the PDF, although {\tt IZI} still
choses the correct higher probability peak as the best-fit solution for this
object. Similarly, the object at the low abundance end (IC2458 -028-007) showing an upper limit in
$Z$ also shows tension between its [SII] flux and the other emission
lines. In this case the low and high abundance solutions in the PDF are merged
into a single broad peak whose low metallicity end hits the $Z_{min}$ value
modeled in the input grids, resulting in an upper limit which is still
consistent with the ``true'' metallicity.

These conflicts raise an important point which we further discuss in
\S7. The results obtained when using this method
will only be reliable as long as the objects being studied are
properly described by the input model grids. That is, if tension exists
between the data and the models, then the best-fit values and
particularly the confidence limits on them can be biased. This is of
great importance to keep in mind when trying to measure the properties
of ionized gas in high redshift galaxies using photo-ionization model
grids aimed at reproducing the observed spectrum of local galaxies and
HII regions where the physical conditions in the gas and the shape of
the ionizing spectrum are not ensured to be similar
\citep[e.g.][]{kewley13b, steidel14}. 

\paragraph{[NII]/[SII] vs [OIII]/H$\beta$} The bottom-left panel of
Figure \ref{fig-5} shows the comparison between this diagnostic and
the fiducial {\tt IZI} metallicities. The agreement is good overall
with zero median offset and a scatter of 0.1 dex. When using
this subset of lines most HII regions show a secondary low probability
peak in the joint PDF towards high metallicities and low ionization
parameters. Sometimes the high abundance end of this peak overlaps
with the $Z_{max}$ value giving rise to the few reported lower
limits. This secondary peak also introduces a high level of asymmetry
in the 1$\sigma$ confidence limits as can be seen in the median
error-bars in Figure \ref{fig-5}. Only in one case (NGC2903 +171+236) this high abundance
solution is chosen instead of the ``true'' value, and this object
stands out as an obvious outlier. Otherwise the deviations observed ar
consistent with the 1$\sigma$ uncertainty. The origin of the
secondary high abundance solutions is related to the absence of the
[OII] doublet in this diagnostic. Leaving out this line
translates into a significantly looser constraint on the ionization
parameter, therefore opening a region of parameter space that would
have been been largely penalized had the [OII] doublet been
included. Since the secondary peak typically has a much lower
probability, we consider $[NII]/[SII]\; vs\; [OIII]/H\beta$ to be a good abundance diagnostic.

\paragraph{[NII]/[SII] vs [OIII]/[SII]} Of the four diagnostics
proposed by \cite{dopita13} this pair of line ratios (bottom-right
panel of Figure \ref{fig-5}) show the highest systematic offset
(-0.08 dex) with respect to the fiducial {\tt IZI} metallicities,
although this deviation is still within the 1$\sigma$ scatter of 0.12
dex and well within the 1$\sigma$ confidence limits in the metallicity
PDF. At high abundances many objects have PDFs that overlap with the
$Z_{max}$ value in the models resulting in several upper
limits. Similarly, two objects have PDFs that overlap with $Z_{min}$
at the low abundance limit. Overall the behavior of this diagnostic
is satisfactory.

\subsection{Classic Literature Diagnostics}

In this section we will evaluate the performance of the six SEL
diagnostics most widely used in the literature: $R23$,
$N2O2$, $N2$, $O3N2$, $O3O2$, and $R3$. 
The top six panels of Figure \ref{fig-6} present these
comparisons. We will find that the performance of some of these diagnostics is extremely
poor unless prior information is given regarding the excitation and ionization
state of the gas. Figure \ref{fig-7} presents a histogram of the
ionization parameter as derived by {\tt IZI} for the V98 sample using the full set of
available SELs. Only objects for which $\log{q}$ is
well constrained by the PDF (93\% of the sample) are
included. The distribution is well fit by a Gaussian distribution with mean
$\mu=\langle \log{q} \rangle=7.29$ and $\sigma=0.23$ dex. This implies
that 71\% (93\%) of the objects in this sample of local HII regions
have ionization parameters that populate a fairly restricted 0.5 dex
(1.0 dex) range around the mean value. 

Many of the diagnostics discussed below have been calibrated
against local samples of HII regions which, like the V98 sample, only
span a very limited dynamical range in ionization and excitation. We
are interested in studying how the performance of different
methods depends on this fact, since this will have important
consequences regarding the use of these diagnostics on objects that are
not ensured to have the same ionization/excitation properties as the calibrators
(e.g. very low metallicity and high redshift galaxies). To obtain
insight on this problem we repeat the comparisons described above
using the best-fit Gaussian distribution shown in Figure \ref{fig-7} as
a prior probability distribution on $\log{q}$ when computing the joint
and marginalized posterior PDFs with {\tt IZI}. The fiducial values
are kept unmodified during this analysis (i.e. calculated without any
priors on $\log{q}$). The result of this experiment is shown in the
bottom six panels of Figure \ref{fig-6}. We now proceed to discuss the
performance of each diagnostic individually:

\begin{figure*}[t]
\begin{center}
\epsscale{1.1}
\plotone{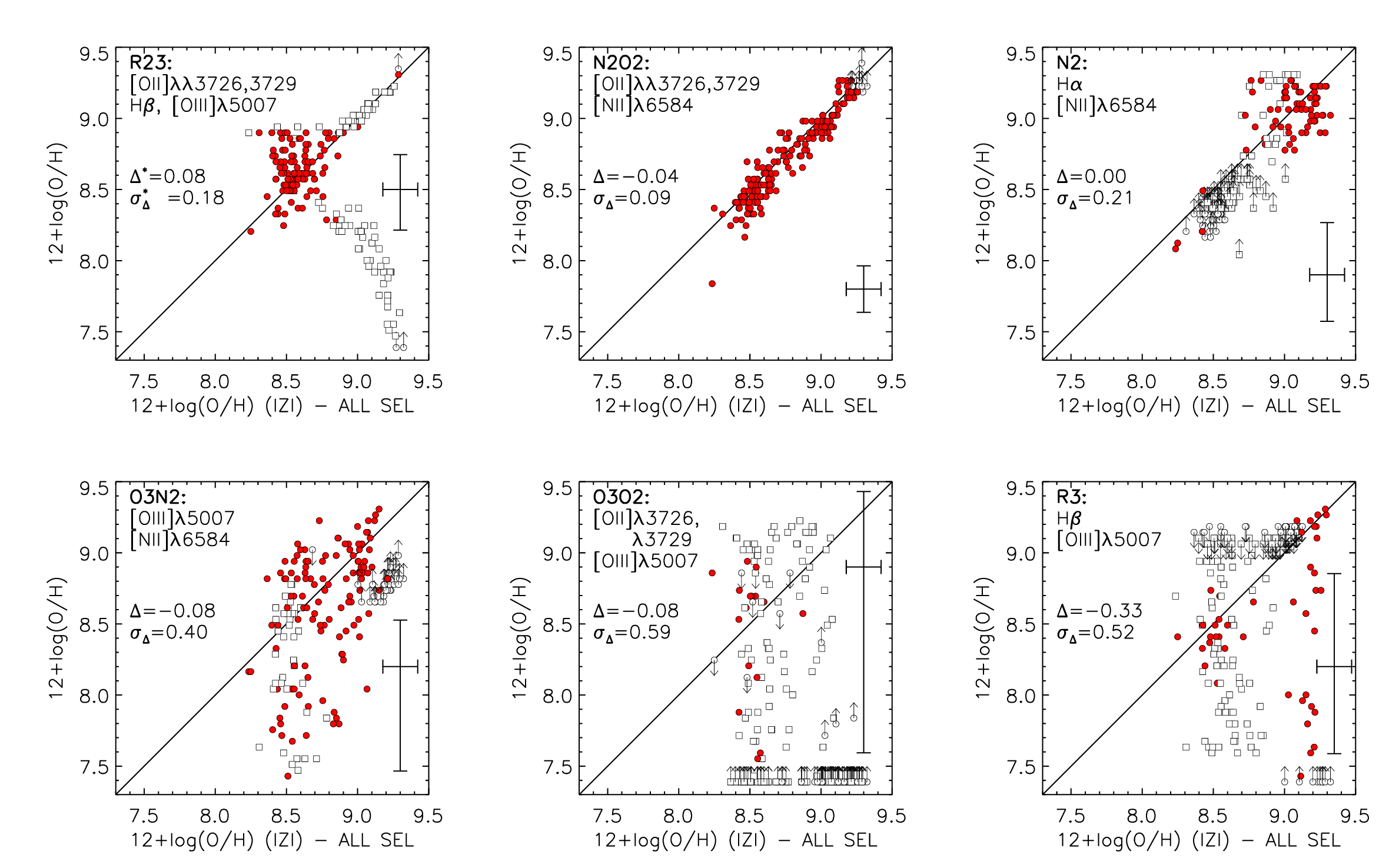}
\rule{0.8\textwidth}{.4pt}
\plotone{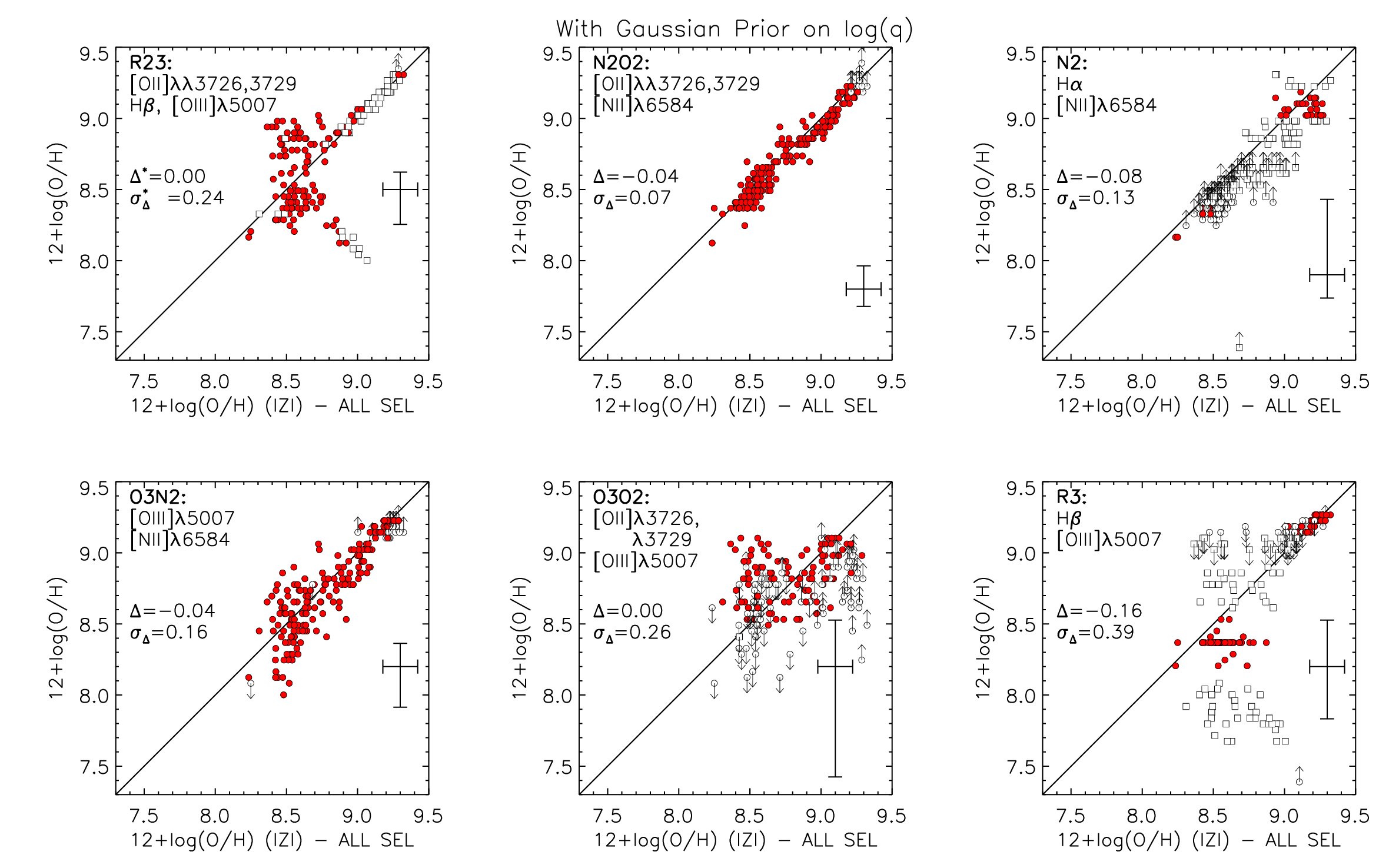}
\caption{{\it Top six panels:} same as in Figure 5 but for subsets of
  lines emulating the $R23$, $N2O2$, $N2$, $O3N2$, $O3O2$, and $R3$
  diagnostics. Maximum ignorance priors on $Z$ and $q$ are assumed in
  these calculations. {\it Bottom six panels:} Same as above but this
  time using the Gaussian prior on the ionization parameter shown in
  Figure \ref{fig-7}. For $R23$ the median offset and scatter are
  calculated only in the $8.3<12+\log{O/H}<8.7$ range.}
\label{fig-6}
\end{center}
\end{figure*}

\paragraph{R23} The double valued nature of the $R23$ diagnostic
already discussed in \S4.1 typically results in a PDF for the
abundance with two probability
peaks corresponding to the so called ``lower branch''  and ``upper branch''
solutions. Most of the HII regions in the V98 sample
are in the upper branch metallicity range of the $R23$ diagnostic. Figure
\ref{fig-6} shows that when only the [OII], [OIII], and H$\beta$
fluxes are known and no prior information is available regarding the
expected metallicity and ionization state of the gas in these objects,
it is not possible to discriminate between the
correct upper branch solution and the incorrect lower branch solution
for about half of the objects at $12+\log{O/H}>8.7$. Below this value
the $R23$ dependance with metallicity flattens out (see Figure
\ref{fig-2}), the two probability peaks in the PDF merge into a single
broad peak, and the scatter increases substantially ($\simeq0.2$ dex
at $8.3<12+\log{O/H}<8.7$). A few objects at the low abundance end of
the sample suffer from the opposite problem with the upper branch
solution being chosen as the best-fit value instead of the correct
lower branch solution. Furthermore, even when the two peaks are merged
together sometimes the PDF is skewed in a way that the mode is shifted
towards higher values, translating into a 0.07 dex offset from the
fiducial value sin the $8.3<12+\log{O/H}<8.7$ range. We do not attempt
to calculate an offset and a scatter at $12+\log{O/H}>8.7$.

Introducing a Gaussian prior on $\log{(q)}$ somewhat alleviates the above
problems as can be seen in Figure \ref{fig-6}. Typically when inspecting the joint posterior PDF one sees
that the upper branch solution lies at $\log{q}$ values that are
higher than those corresponding to the lower branch solution. At the
high abundance end, these values are closer to the mean of the $\log{q}$ distribution presented in
Figure \ref{fig-7}. The second row of panels in Figure \ref{fig-1}
provides a good example of this behavior. Therefore using the Gaussian
prior on $\log{q}$ tends to decrease the probability of the lower branch
solution for objects in the high metallicity end of the sample,
bringing the $R23$ abundance of many objects into agreement
with the ``true'' abundance. The opposite effect happens at the low
abundance end, where the ionization parameters associated with the
lower branch solution are typically closer to the mean of the
$\log{q}$ distribution, so objects for which the $R23$ diagnostic was
choosing the wrong upper branch solution when no prior on $\log{q}$
was used also come into better agreement with the fiducial
values. This lowers the systematic offset in the
$8.3<12+\log{O/H}<8.7$ range from 0.07 dex to -0.03 dex, but is
accompanied with an increase in the scatter in this region from
0.18 dex to 0.24 dex. Regardless of these improvements, the wrong solution is still
preferred for a significant fraction of the objects. 

In line with what is common usage in the literature
\citep[e.g.][]{pilyugin05, kewley08, moustakas10}, we conclude that
$R23$ can only be used as a metallicity diagnostic if prior
information regarding the expected metallicity range (i.e. the branch)
is available. In order to provide a fair evaluation of the performance
of this diagnostic we include an extra prior on the metallicity
that assigns all objects in the V98 sample to the upper branch. This
prior is a step function with uniform probability per decade in $Z$ above
$12+\log{O/H}=8.3$ and zero probability below this value. The results
of using this branch prior and of using both the branch and Gaussian
ionization parameter priors together are show in the top left  and
bottom left panels of Figure \ref{fig-8} respectively. After including prior information
on the expected metallicity range the performance of
the $R23$ diagnostic is excellent at the high abundance end
($12+\log{O/H}>8.8$), but degrades dramatically below this value where
ratio becomes highly insensitive to the metallicity. The scatter in
the $8.3<12+\log{O/H}<8.7$ remains high at 0.24 dex, and we measure a
global scatter of 0.2 dex over the full metallicity range of the V98
sample. Introduction of the prior information on the branch removes any
systematic offset with respect to the fiducial abundances. 

\begin{figure}[t]
\begin{center}
\epsscale{1.2}
\plotone{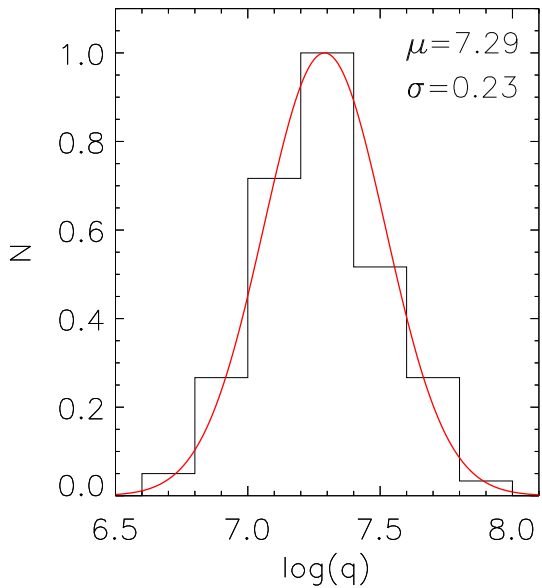}
\caption{The black histogram presents the normalized distribution of
  best-fit ionization parameters for the V98 sample using all
  available SELs. Only sources where the marginalized PDF is single
  peaked and bounded on both sides are included (93\% of the
  sample). The best-fit Gaussian PDF adopted as a prior on $\log{(q)}$
in the bottom six panels of Figure \ref{fig-6} and the bottom three
panels of Figure \ref{fig-8} is shown in red. Also reported are the
best-fit parameters of this Gaussian PDF.}
\label{fig-7}
\end{center}
\end{figure}

\paragraph{N2O2} As can be seen in Figure \ref{fig-6} this diagnostic
performs extremely well over the full metallicity range analyzed
here. It shows a median offset of -0.04 dex and a scatter of
0.09 dex with respect to the fiducial values. For a few objects at the high abundance
end the PDF overlaps with the $Z_{max}$ limit of the input grids
resulting in a few lower limits. In \S4.1 we discussed how part of the
power of $N2O2$ as a metallicity diagnostic comes from the fact that
this line ratio is very insensitive to the ionization parameter. This
is evident from the fact that the results remain almost unchanged when using a Gaussian
prior on the $\log{(q)}$ distribution. 

At first sight our results seem
to imply that reliable abundances can be
obtained from this line ratio without any prior knowledge of the
expected metallicity range nor the ionization and excitation
conditions in the gas. This statement is subject to a very important
caveat already mentioned in \S4.1. As shown in \cite{perez-montero14}
the $N2O2$ diagnostic traces the N/O
abundance almost linearly. The photo-ionization models used here
assume a monotonic single valued function between N/O and O/H with no
scatter, and the ability of $N2O2$ to trace the oxygen abundance
relies heavily on this assumption. In reality the observed correlation
between these two quantities shows a scatter of 0.15-0.25 dex and the
N/O ratio appears to increase systematically as a function of
galaxy total stellar mass \citep[e.g][]{vanzee98a, andrews13,
  perez-montero14, belfiore14}. In the absence of systematic trends
like this latter one we could simply incorporate the scatter in the N/O vs
O/H relation to the error-bars on the abundance calculated by {\tt
  IZI}. Since both are of a similar magnitude we do not expect the
issue of the scatter to be of great significance. The most worrisome
problem is the presence of systematic trends in the N/O vs O/H relation as a
function of global and local galactic properties and the increased
scatter in N/O towards the low abundance regime. The proper way to
deal with these systematics is to leave N/O as a free parameter in the
fits as done by \cite{perez-montero14}. In the future we plan to
extend the capabilities of {\tt IZI} to deal with extra parameters
beyond $Z$ and $q$ in order to deal with these type of problems.

\begin{figure*}[t]
\begin{center}
\epsscale{1.1}
\plotone{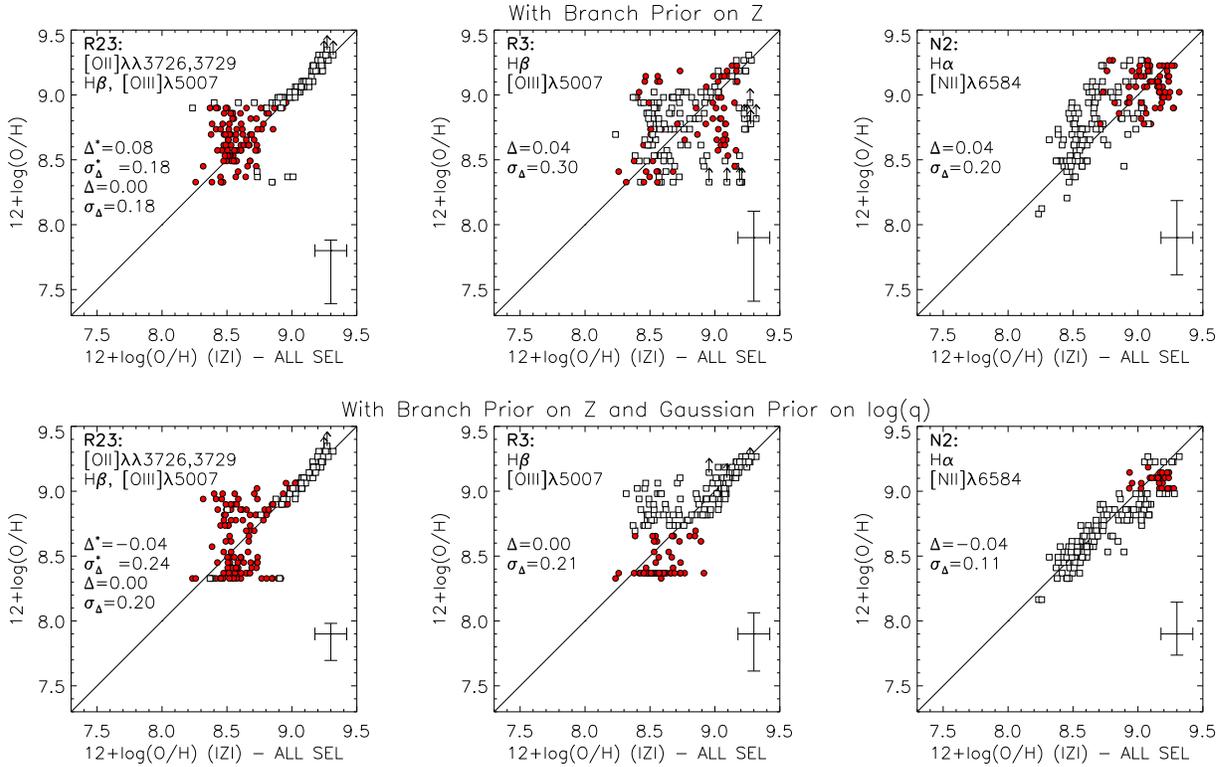}
\caption{Same as in Figure 6 for $R23$, $R3$, and $N2$ but now using a
branch prior in the top three panels, and and both a branch prior and
a Gaussian ionization parameter prior in the bottom three panels. For
$R23$ we now report the offset and scatter in both the
$8.3<12+\log{O/H}<8.7$ range ($\Delta^*$, $\sigma_{\Delta}^*$) and the
full range of the data ($\Delta$, $\sigma_{\Delta}$).}
\label{fig-8}
\end{center}
\end{figure*}

\paragraph{N2} Figure \ref{fig-6} shows a large number of objects with
lower limits and double peaked PDFs in metallicity when using the $N2$
diagnostic. The reason behind this behavior is the double valued
nature of this diagnostic (see Figure \ref{fig-2} and the bottom
panels of Figure \ref{fig-1}) which is rarely appreciated in the
literature \citep[although see][]{kewley02} because the flattening and
inversion of the metallicity dependance of
$N2$ takes place at relatively high abundances
($12+\log{O/H}\simeq9.2$). In the sample used by
\cite{pettini04} to calibrate this method there are only a couple
objects at $12+\log{O/H}\simeq9.2$ so the ``upper branch'' of $N2$ has
not been exposed observationally. A closer inspection of Figure 1 in
\cite{pettini04} actually shows evidence for this plateau in $N2$ at the high
abundance end. The \cite{dopita13} photoionization models sample
the oxygen abundance past the region where this inversion happens
(upper right panel of our Figure \ref{fig-2}).

With this in mind we can better understand the results presented in
Figure \ref{fig-6}. For objects showing single valued PDFs (red circles)
the entirety of the high abundance peak (i.e. the upper branch
solution) is past the $Z_{max}$ value in the models. Most objects at
lower abundances show two peaks in the PDF corresponding to the lower
and upper branch solutions for $N2$. For the V98 sample the low
abundance peak is typically the ``correct'' solution, but
without any prior information available on either the abundance nor the
ionization parameter many times the wrong high abundance solution
is preferred. This is happens most often in the $8.7<12+\log{O/H}<9.1$
range. Towards lower abundances upper limits become common as the two
peaks become more widely separated and the upper branch solution
overlaps with the $Z_{max}$ value. Overall, ignoring lower limits this
diagnostics shows zero systematic deviation from the fiducial values
and a large scatter of 0.21 dex.

Including a Gaussian prior on the ionization parameter censors regions of parameter
space where the ionization parameter is high and this reduces the
number of objects for which the high abundance solution is
preferred. This is because the two parameters are correlated with high
abundance solutions typically showing higher values of $q$. The
scatter in this case is significantly reduced to 0.13 dex, but the
median offset increases to -0.08 dex. In any case, the number of
objects for which we can only provide lower limits in the metallicity
remains high.

The double valued nature of $N2$ implies the need of choosing a
branch, just as it is typically done when using the $R23$ diagnostic
(see discussion above). In the case of the V98 sample all objects
populate lower branch. In the top right panel of Figure \ref{fig-8} we
show the effects of imposing a prior on the metallicity in order to emulate
this preference for the lower branch solutions. We use a flat prior
with equal probability per decade in $Z$ at $12+\log{O/H}<9.3$ and
zero probability above this value. This decreases the median offset
but the scatter remains large ($0.2$ dex) because of the covariance
between the abundance and the ionization parameter. We find that only after introducing
prior information on both the metallicity range (i.e. the branch) and
the ionization parameter of the gas $N2$ becomes a
well behaved abundance diagnostic with a median offset of -0.04 dex
and scatter of 0.11 dex. Finally, just as $N2O2$, $N2$ is also affected by the assumed
behavior of the N abundance discussed above \citep{perez-montero14}.

\paragraph{O3N2} This diagnostic, first proposed by \cite{alloin79}
and latter recalibrated by \cite{pettini04} shows a very poor performance with a
median offset of -0.08 dex with respect to the fiducial metallicities
and a large scatter of 0.4 dex. This is not surprising after
inspection of Figure \ref{fig-2}, which shows
that $O3N2$ is as sensitive to the ionization parameter as it is to
the metallicity. 

A scatter of 0.4 dex is significantly higher than the
0.25 dex scatter reported by \cite{pettini04} in their
calibration. The cause behind this discrepancy is the limited dynamical range in
ionization parameter spanned by the calibrators. To show this, Figure
\ref{fig-6} also presents the results of using the
Gaussian prior in $\log{q}$ discussed above in our calculations. This
dramatically improves the performance of the $O3N2$ diagnostic by
lowering the median offset to -0.04 dex and the scatter to 0.16 dex,
in much better agreement with \cite{pettini04}. This implies that
$O3N2$ can only be used as a reliable abundance estimator if prior
information regarding the ionization and excitation state of the gas
is available. This diagnostic is also strongly affected by the N
abundance \citep{perez-montero14}.

\paragraph{O3O2 and R3} Figure \ref{fig-6} shows that these two ratios perform very poorly as
abundance diagnostics with median offsets of -0.08 dex and -0.33 dex,
and scatters of 0.59 dex and 0.52 dex (i.e. factors of 3) for $O3O2$
and $R3$ respectively. This is not surprising in light of the strong
dependance on the ionization parameter seen in Figure \ref{fig-2} for
these diagnostics. In fact, both line ratios have been typically used as
ionization parameter diagnostics in the literature
\citep[e.g.][]{mcgaugh91, kewley02, lilly03, nakajima13}.
\cite{maiolino08} combined a sample of low
metallicity galaxies with direct $T_e$ abundances with a sample of
high metallicity star forming galaxies from SDSS-DR4 with
metallicities measured using the \cite{kewley02} photo-ionization
model prescriptions, to calibrate a series of SEL diagnostics over a
broad dynamical range metallicity. In particular, they find
relatively tight correlations with metallicity for $O3O2$ and $R3$
with typical scatter of $\sim0.1-0.3$ dex, so
the authors adopt them as part of their suite of abundance diagnostics. 

The apparent contradiction between the observed correlation of
these line ratios with metallicity in \cite{maiolino08} and the poor performance seen in
Figure \ref{fig-6} can again be explained by the limited dynamical
range in ionization and excitation conditions in the calibration
sample. Similar to what we observed for $O3N2$,
the performance of these two diagnostics improves significantly when
using our Gaussian prior on $\log{(q)}$. The bottom panels of Figure
\ref{fig-6} show that this reduces the median offsets to
zero and -0.16 dex, and the scatter to 0.26 and 0.39 dex for $O3O2$
and $R3$ respectively, closer to the scatters seen in the
\cite{maiolino08} calibrations. Even after these improvements the
performance of $O3O2$ and $R3$ as abundance diagnostics is still poor. 

In the case of $R3$, Figure \ref{fig-2} shows that it follows a double
valued behavior similar to that of $R23$ (although with a much
stronger dependance on the ionization parameter) so proper usage of this
diagnostic also requires choosing between a lower and upper branch. The
middle panels of Figure \ref{fig-8} present the results
of using a branch prior on the metallicity with uniform probability per decade in $Z$ above
$12+\log{O/H}=8.3$ and zero probability below this value, and a
combination of a branch prior and a Gaussian prior on $\log{(q)}$. The
branch prior marginally improves the performance of this diagnostic
and it is only after including priors on both $q$ and $Z$ that its
performance becomes acceptable at the high abundance end but
degrades rapidly at $12+\log{O/H}<8.9$ as the ratio becomes
insensitive to the metallicity. 

Overall, we do not consider $O3O2$ to be a reliable abundance
indicator and we recommend against using $R3$ without previous
knowledge of the ionization conditions and expected metallicity range
of the objects under study. 

\subsubsection{Classic SEL Diagnostics Performance Summary}

In summary, out of the six SEL diagnostics examined above only $N2O2$
seems to provide reasonably good results without requiring prior knowledge
of the ionization and excitation conditions in the gas and/or the
expected range in metallicity of the objects. This diagnostic also presents the
lowest scatter among the six methods, but it is subject to systematic
uncertainties associated with the assumed behavior of the N/O
abundance as a function of metallicity. Furthermore, these two lines
are well separated in wavelength and using them requires an accurate
correction for dust attenuation and enough spectral resolution to
separate [NII]$\lambda$6583 from H$\alpha$.

Our analysis shows that reliable results, although with an increased
level of scatter can also be obtained from $N2$ and $O3N2$ if prior
information is available regarding the ionization conditions in the
gas. Furthermore, the performance of the $N2$ diagnostic can become
comparable to that of $N2O2$ if on top of having prior information on
$q$, we know in advance in which branch
objects will fall. This is typically the case with most samples of
galaxies and HII regions since the separation between the lower and
upper branches of $N2$ happens at a very high abundance.
The separation between the lower and upper branches of the $R23$
diagnostic on the other hand occurs at a much lower
abundance, so $R23$ can only be used reliably if the branch is known
a priori, and even in that case the scatter becomes uncomfortably
large at intermediate values where the ratios are insensitive to the
metallicity \citep[see also][]{lopez-sanchez12}. 

Finally, we find that the $O3O2$ and $R3$ ratios are always fairly
poor abundance diagnostics. The results go from catastrophic to poor
when prior information on the ionization parameter is included. For
$R3$ the results become marginally acceptable if further information
regarding the branch in which the objects fall is available. 

The analysis presented above is conducted in the context of
the photo-ionization models adopted as input for {\tt IZI}. While
these models surely do not provide perfect representations of real star
forming galaxies and HII regions, general trends of relative SEL
intensity as a function of metallicity and ionization parameter like
those presented in Figure \ref{fig-2} are thought to be robust, at least in a
qualitative sense. Therefore, while the quantitative details of the
analysis might change as a function of the input model adopted, the
conclusions presented in this section should not be significantly
affected, at least in a qualitative way.

We emphasize here that the above analysis is aimed at evaluating the
intrinsic performance of these diagnostics in terms of how much
information useful to constrain the metallicity is encoded in
these line ratios. So far we have not evaluated any systematics associated
with the calibration of these diagnostics against either direct method
abundances or photo-ionization models. This is the subject of the next section.

\section{Comparing Photo-ionization Model, Direct Method, and Recombination Line Abundances}

In this section we explore the discrepancies in the abundance scale
known to exist between the different families of methods used to
measure chemical abundances in ionized gas. In particular we compare
oxygen abundances measured with {\tt IZI} 
using different photo-ionization models in the literature with those obtained by
applying the direct method, a set of empirically calibrated SEL
methods (i.e. calibrated against direct method abundances), and abundances
derived from the direct observation of oxygen recombination lines (RL).

Unlike CELs and temperature sensitive auroral lines, the emissivity
of recombination lines has a very mild dependance on $T_e$ and
therefore on the presence of temperature fluctuations within
nebulae. This method does not suffer from the systematic uncertainties
that affect the direct and SEL methods, although it is subject to its own set of systematic
uncertainties chiefly having to do with the accurate calculation of
recombination coefficients to high quantum level states \citep{liu00,
stasinska04}. Furthermore, RL abundances have been shown to
agree very well with the stellar abundances of OB stars in star
forming regions \citep[e.g.][]{peimbert05, simon-diaz11}. The above makes them a
good reference point to evaluate the performance of other methods.

Comparisons of this type have been performed before by
\cite{garcia-rojas07} and \cite{lopez-sanchez12} for a handful of
local HII regions with RL measurements. These studies find
that in HII regions the direct method consistently yields abundances that are
$\sim0.2$ dex lower than the RL method. \cite{garcia-rojas07}
conclude that this offset is best explained
by the presence of temperature fluctuations inside HII regions, which
tend to bias the direct $T_e$ measurements towards higher values
and the abundances towards lower values, although other authors
disagree \citep[c.f.][]{simon-diaz11}. \cite{lopez-sanchez12} compare
the average abundances from several empirically calibrated and theoretically
calibrated SEL diagnostics to RL abundances and concludes that SEL
methods calibrated against photo-ionization models yield abundances
that are consistently $\sim0.3$ dex higher than the direct method and
$\sim0.1$ dex higher than the RL method. This is a similar
difference to the one seen between empirical and theoretical SEL
diagnostics by \cite{kewley08} using a large sample of
emission line galaxies from SDSS.

{\tt IZI} allows us to calculate the oxygen abundance of HII regions
using an arbitrary photo-ionization model and without having to restrict
the analysis to the choice of a particular set of SEL diagnostics. This offers
the opportunity to extend the above comparisons to test how
the results from different photo-ionization models compare to direct
and RL abundances, independently from the systematics associated with a
particular calibration of a particular diagnostic.

As mentioned in \S1 the recombination lines of elements heavier than He
are typically $10^{3-4}$ times fainter than SELs, so measuring them in
local HII regions typically requires several hours of exposure time
using high resolution spectrographs in 10 meter class
telescopes. Because of this, RL abundances have only
been measured for He, C, O, and Ne in a few dozen bright HII regions
in the MW and the Local Group \citep[e.g.][]{peimbert03, 
  peimbert05, tsamis03, esteban04, esteban09, garcia-rojas07,
  lopez-sanchez07, bresolin09}. For this comparison we use the compilation of 22 local HII
regions with measured RL oxygen abundances presented in Table 6\footnote{Because of an error during the
    creation of this table, the line ratios reported in
    \cite{lopez-sanchez12} are incorrect (Angel R. Lopez-Sanchez
    private communication). We used the line fluxes and errors
    as reported in the original reference for each object.} of
  \cite{lopez-sanchez12}. Although hundreds of emission lines are
  typically detected in the high quality spectra used to measure RL
  abundances in these regions, we run IZI using only the [OII]$\lambda\lambda$3726,3729, H$\beta$,
  [OIII]$\lambda$5007, H$\alpha$, [NII]$\lambda$6583, and
  [SII]$\lambda\lambda$6717,6731 line fluxes in order to emulate the
  information that would typically be available from a low S/N, low
  resolution spectrum.

\begin{figure*}[t]
\begin{center}
\epsscale{0.8}
\plotone{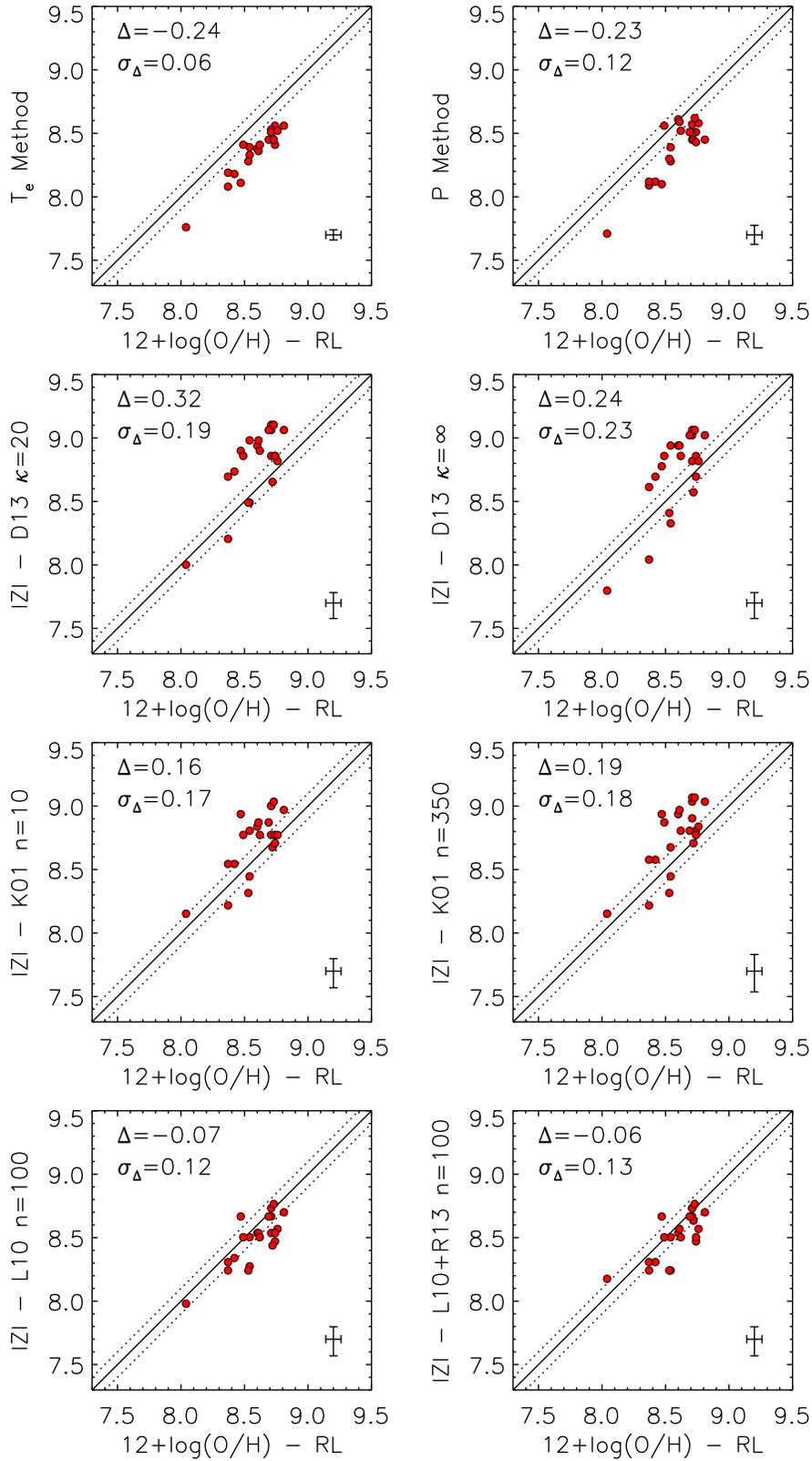}
\caption{Oxygen recombination lines (RL) abundances versus those derived using different
  methods. RL abundances are plotted in the x-axis and
  together with $T_e$ method and $P$ method abundances (first
  row) are taken from Table 6 of \cite{lopez-sanchez12}. The
  y-axis in the rest of the panels show abundances computed with IZI
  using the [OII]$\lambda\lambda$3726,3729, H$\beta$,
  [OIII]$\lambda$5007, H$\alpha$, [NII]$\lambda$6583, and
  [SII]$\lambda\lambda$6717,6731 line fluxes from the original
  references in Table 6 of \cite{lopez-sanchez12} and different
  photo-ionization model grids. The second row corresponds to the
  \cite{dopita13} models for $\kappa=20$ and $\kappa=\infty$
  (i.e. Maxwell-Boltzmann $T_e$ distribution). The third row
  corresponds to the \cite{kewley01} models with $n=10$ and
  $n=350$. The bottom row shows results for the \cite{levesque10}
  models on the left and the combined \cite{levesque10} and
  \cite{richardson13} models with extended sampling towards high value
  sof $q$ on the right. Dotted lines at $\pm0.1$ dex offsets are shown
  for reference.}
\label{fig-9}
\end{center}
\end{figure*}

Figure \ref{fig-9} presents the results of the comparison. The first row of
panels compares the RL abundances of the 22 regions against the
direct method abundances (left) computed using the same high S/N, high
resolution spectra used to derive the RL abundances, and against the
average of three empirically calibrated SEL methods from
\cite{pilyugin01}, \cite{pilyugin05}, and \cite{pilyugin10} (dubbed
the $P$-method). The
data presented in these two panels is taken from
\cite{lopez-sanchez12}. As already stated in that work, both direct
method abundances and empirically calibrated SEL abundances
consistently under-predict the RL abundances by 0.24 dex and 0.23 dex
respectively. Direct method abundances show a scatter of 0.06 dex
with respect to RL abundances. This is consistent with the errors in
the direct and RL abundance determinations which have a median value of 0.04
dex and 0.06 dex respectively. On the other hand, the empirically
calibrated SEL abundances show a scatter of 0.12 dex against the RL
abundances. This is significantly larger than the reported scatter of
0.075 dex for the \cite{pilyugin10} calibrations.

The rest of the panels in Figure \ref{fig-9} compare abundances
derived with IZI using different photo-ionization models against
RL abundances. We present results for a subset of the photo-ionization
models presented in \cite{kewley01}, \cite{levesque10},
\cite{richardson13}, and
\cite{dopita13}. The models have been described in \S3.2. The
\cite{dopita13} models systematically over-predict the RL abundances
by 0.32 dex and 0.24 dex for electron energy kappa probability
distributions with $\kappa=20$ and a $\kappa=\infty$
(i.e. a Maxwell-Boltzmann distribution) respectively. The scatter seen
for these two models is 0.19 dex and 0.23 dex respectively. Similarly
to the case of the $P$-method above, the observed
scatter is significantly larger than what is expected from the median
of the uncertainty calculated by IZI which corresponds to 0.1
dex\footnote{This median uncertainties corresponds to the median between
the upper and lower error-bars for both versions of the models presented.}. Although
the dynamical range in abundance of the HI region sample with RL line
measurements is limited, and the number of regions is small, there
appears to be a trend of an increasing overestimation towards higher
abundances.

 Results for the \cite{kewley01} models are presented in the third row
of Figure \ref{fig-9} assuming two
different values for the electron density ($n_e=10$ cm$^{-3}$ and $n_e=350$
cm$^{-3}$). Abundances derived using these models are higher
than RL abundances with offsets of 0.16 dex and
0.19 dex and scatters of 0.17 dex and 0.18 dex for the $n_e=10$ cm$^{-3}$ and $n_e=350$
cm$^{-3}$ models respectively.  No significant difference is seen between the
two electron density regimes considered. The observed scatter is slightly larger
than the 0.13 median uncertainty reported by IZI. 
This offset is similar to the offset measured by
\cite{lopez-sanchez12} when comparing the average of a few
theoretically calibrated SEL diagnostics based on the \cite{kewley01}
models to RL measurements. The observed offset is similar in magnitude
to the offset measured for direct abundance based methods but in the
oposite direction. A similar trend to that observed for the
\cite{dopita13} models is seen in which the difference in the values
become larger towards the high abundance end.

Results from the \cite{levesque10} models both with and without
including their extension towards high values of $q$ by
\cite{richardson13} are presented in the two bottom panels of Figure
\ref{fig-9}. The inclusion of the \cite{richardson13} models leaves
results practically unchanged since none of the HII regions in this
sample presents extremely large ionization parameters. Abundances
derived using the \cite{levesque10} models present the best agreement
with RL abundances among all the photo-ionization models considered
here. This model shows a systematic offset of -0.07 dex and a scatter of 0.12
which is in excellent agreement with the median error of 0.11 dex
reported by IZI. This is the best performance both in terms of the
median offsets and the magnitude of intrinsic scatter beyond the
measurement errors among all the SEL methods studied here.

This comparison highlights an interesting fact. Abundances derived
using IZI in combination with different photo-ionization models show
different offsets with respect to abundances measured using the direct
and RL methods. A large range in median offsets is seen among
different photo-ionization models, although this is driven mainly by
the \cite{dopita13} models which predict abundances typically $\sim0.1$ dex
higher than the \cite{kewley01} and $\sim0.3$ dex higher than the
\cite{levesque10} models. The MAPPINGS photo-ionization code used to
compute all these models suffered a major revision between the latter two works, and the
former. These upgrades are thoroughly discussed in \cite{dopita13} and
include among other things the inclusion of a kappa function to
describe the electron energy distribution instead of the classically
used Maxwell-Boltzmann distribution \citep{nicholls12}. Our results
suggest that this is not the main cause behind the observed
discrepancies, as adopting a Maxwell-Boltzmann distribution
($\kappa=\infty$) only decreases the offset by less than 0.1 dex.
Other factors like the assumed nebular geometry, the updated atomic
data and abundance set and the different parametrization of the N/O vs
O/H relation must contribute to the observed offsets. This also
implies that our results do not contradict the fact that a $\kappa$
electron energy distribution can help reconcile direct and RL
abundances as proposed by \cite{nicholls13}, as the systematic offets
observed in this work are most likely not associated with the adoption
of the $\kappa$ distribution. In the future we
expect to explore in more detail the effects on the derived abundances
caused different assumptions in photo-ionization models.

From this analysis we can conclude that Bayesian inference using
photo-ionization models can be successful at reproducing the results
of the RL method. We do not find a significant discrepancy between RL
abundances and the abundances computed using IZI with the
\cite{levesque10} models. The large offset seen against the
\cite{dopita13} models highlights the presence
of poorly understood systematic uncertainties affecting
photo-ionization models. \\

\section{Summary and Conclusions}

We have presented a new method to measure the metal abundance and
ionization parameter of HII regions and star forming galaxies using
SELs. The method is based on the application of Bayesian inference to
calculate the joint and marginalized PDFs for these two parameters
given an arbitrary set of emission line flux measurements or upper
limits and a model for how the brightness of these lines depend on $Z$
and $q$. We also present an implementation of the method called {\tt
  IZI} that computes these PDFs using theoretical photo-ionization
models and we make the code publicly.

Using {\tt IZI} we have tested the performance of a series of popular
SEL diagnostics in the literature to evaluate how much information
regarding the metal abundance is carried by these line ratios. For
this we have used a sample of 186 extragalactic HII regions from V98
and have run {\tt IZI} using subsets of emission lines that emulate
the information contained in a particular diagnostic. We have
evaluated the $R23$, $N2O2$, $N2$, $O3N2$, $O3O2$, and $R3$ line
ratios, as well as the four pairs of line ratio diagnostics proposed
by \cite{dopita13}.

We have also used a sample of bright local HII regions with direct
method and RL abundance measurements to study the discrepancies in
the abundance scale between these two methods and those based on the
use of theoretical photo-ionization models. For this comparison we
include results based on using {\tt IZI} in combination with several
photo-ionization models in the literature including those presented in
\cite{kewley01}, \cite{levesque10}, \cite{richardson13}, and
\cite{dopita13}.

From the experiments presented above we conclude the following:

\begin{itemize}

\item Bayesian inference provides an optimal tool to measure the
  physical conditions of ionized gas in nebulae. Our method has many
  advantages over the classic approach of calibrating a single line
  ratio or a pair of line ratios as an abundance diagnostic. These
  advantages are listed in \S1. Here we highlight the fact that {\tt
    IZI} uses all the available spectroscopic information
  simultaneously, circumvents the need to calibrate a particular
  diagnostic against a particular set of models, and allows one to
  naturally identify multiple peaks, asymmetries, and upper or lower
  limits in the PDF which translates into realistic error estimates.

\item When evaluating the performance of different SEL abundance
  diagnostics in the literature we find a large diversity in terms of
 how accurately the diagnostics trace abundance of HII regions. Some
 diagnostics only perform well when prior information is available
 regarding the abundance range (i.e. the branch) in which the
 sources are expected to fall, or the ionization and excitation
 conditions in the gas. The $N2O2$ diagnostic seems to be robust
 against the lack of prior information, but its dependance with the
 N/O ratio make it subject to significant systematic
 uncertainties. In light of their poor performance we recommend
 strongly against the use of $R3$ and  $O3O2$ as abundance
 diagnostics. Other diagnostics like $R23$ and $N2$ can only perform
 well if prior information regarding the correct branch is
 available. Finally $O3N2$ only performs reasonably well if prior
 information regarding $q$ is available. The four pairs of SEL ratios
 proposed by \cite{dopita13} carry enough information to constrain $Z$
 and $q$ simultanously in most cases. 

\item The different limitations of individual diagnostics highlight
  the advantage of using all spectroscopic information available from
  transitions of different elements and different ions when computing
  abundances, since this minimizes systematics and helps constrain
  both the abundance and the ionization parameter in a much better way
  than individual diagnostics. 

\item Abundances derived through Bayesian inference using
  photo-ionization models can agree very well with RL abundances, but
  the agreement depends on which models are used. Out of all the
  models tested here the \cite{levesque10} grids produce abundances
  that agree with RL measurements remarkably well (to within
  30\%). The models of \cite{kewley01} show a $0.15-0.2$ dex
  systematic offset towards higher abundances but this offset is still
  within the observed scatter. The \cite{dopita13} models significantly
  overpredict RL abundances by 0.2-0.3 dex for this small sample of
  local HII regions, although the discrepancy is not associated with
  the use of a $\kappa$ electron energy distribution. The scatter seen
  between different photo-ionization models highlight the
  systematic uncertainties that affect them, including the assumptions
  regarding the input ionizing flux from synthetic stellar population
  models, the assumed behavior of relative elemental abundances like
  N/O vs O/H, and the assumed physical structure of the nebulae.

\item Direct method abundances are found to be systematically lower
  than both RL and photo-ionization model abundances. This is in
  agreement with previous claims in the literature and is most likely
  caused by the unaccounted effect of temperature fluctuations within
  ionized nebulae.

\end{itemize}
 
Finally we would like to stress two important points regarding the use of {\tt
  IZI}. The first is the fact that the results produced by {\tt IZI} will
only be reliable if the input model being used corresponds to a proper
characterization of the sources being analyzed. If the shape of the input ionizing
spectrum or the assumed pattern of relative elemental abundances in the models are
significantly different from what they really are in the objects under
study, then even if a correct statistical analysis is used to
compare the data to these models, results will be subject to
systematic uncertainties. Second, as briefly mentioned in \S3.2,
this method is flexible and not limited to be used with theoretical
photo-ionization models. In principle the user can adopt any model that
traces the behavior of the lines of interest as a function of $Z$ and
$q$, even one based on empirical grids of direct method abundances for
real objects. In the future we expect to explore these possibilities
and also to expand {\tt IZI} to deal with extra parameters like
the N/O abundance, the electron density, and the age or effective
temperature of the ionizing stellar population.\\

We would like to thank Gwen Rudie, Andrew Benson, Daniel Masters, and
Juna Kollmeier for useful discussions regarding this paper. This
research made use of NASA's Astrophysics Data System, the
NASA/IPAC Extragalactic Database (NED) which
is operated by the Jet Propulsion Laboratory, California Institute of
Technology, under contract with the National Aeronautics and Space
Administration, and the VizieR catalogue access tool, CDS, Strasbourg,
France. The original description of the VizieR service was published
in A\&AS 143, 23.

\end{document}